\documentclass{article}
\usepackage{ijcai25}

\usepackage{times}
\usepackage{soul}
\usepackage{url}
\usepackage[hidelinks]{hyperref}
\usepackage[utf8]{inputenc}
\usepackage[small]{caption}
\usepackage{graphicx}
\usepackage{amsmath}
\usepackage{amsthm}
\usepackage{booktabs}
\usepackage{algorithm}
\usepackage{algorithmic}
\usepackage[switch]{lineno}
\usepackage{float}  

\graphicspath{{images/} }
\usepackage{bbold}
\newtheorem{theorem}{Theorem}[section] 
\usepackage{amssymb}
\usepackage{multirow}
\usepackage{bbding}
\usepackage{subfigure}
\usepackage{arydshln}
\usepackage{setspace}
\usepackage{tcolorbox}
\usepackage{algorithm}
\usepackage{algorithmic}

\title{CueGCL: Cluster-aware Personalized Self-Training for Unsupervised Graph Contrastive Learning}

\author{
Yuecheng Li$^1$\and
Lele Fu$^2$\and
Sheng Huang$^{2}$\and
Chuan Chen$^1$\and
Lei Yang$^{1}$\and
Zibin Zheng$^3$\\
\affiliations
$^1$School of Computer Science and Engineering, Sun Yat-sen University, Guangzhou 510275, China\\
$^2$School of Systems Science and Engineering, Sun Yat-sen University, Guangzhou 510275, China\\
$^3$School of Software Engineering, Sun Yat-Sen University, Zhuhai, 519082, China\\
\emails
\{liych78, fulle, huangsh253\}@mail2.sysu.edu.cn,
\{chenchuan, yanglei39, zhzibin\}@mail.sysu.edu.cn,
}

\begin{document}

\maketitle

\begin{abstract}
    Recently, graph contrastive learning (GCL) has emerged as one of the optimal solutions for node-level and supervised tasks. However, for structure-related and unsupervised tasks such as graph clustering, current GCL algorithms face difficulties acquiring the necessary cluster-level information, resulting in poor performance. In addition, general unsupervised GCL improves the performance of downstream tasks by increasing the number of negative samples, which leads to severe class collision and unfairness of graph clustering. To address the above issues, we propose a \textbf{\underline{C}}l\textbf{\underline{u}}ster-awar\textbf{\underline{e}} \textbf{\underline{G}}raph \textbf{\underline{C}}ontrastive \textbf{\underline{L}}earning Framework (\textbf{CueGCL}) to jointly learn clustering results and node representations. Specifically, we design a \textit{\underline{pe}rsonalized \underline{s}elf-\underline{t}raining (PeST)} strategy for unsupervised scenarios, which enables our model to capture precise cluster-level personalized information. With the benefit of the PeST, we alleviate class collision and unfairness without sacrificing the overall model performance. Furthermore, \textit{\underline{a}ligned \underline{g}raph \underline{c}lustering (AGC)} is employed to obtain the cluster partition, where we align the clustering space of our downstream task with that in PeST to achieve more consistent node embeddings. Finally, we theoretically demonstrate the effectiveness of our model, showing it yields an embedding space with a significantly discernible cluster structure. Extensive experimental results also show our CueGCL exhibits state-of-the-art performance on five benchmark datasets with different scales.
\end{abstract}

\section{Introduction}

With the development of the information age, data with complicated relationships have emerged in large quantities, such as social networks. Graph clustering (Community detection) \cite{li2024CDNMF,deng2024thesaurus} is an essential task in complex network analysis.

\begin{figure*}[t]
\centering
\subfigure[The raw feature]{
\includegraphics[width=0.31\textwidth]{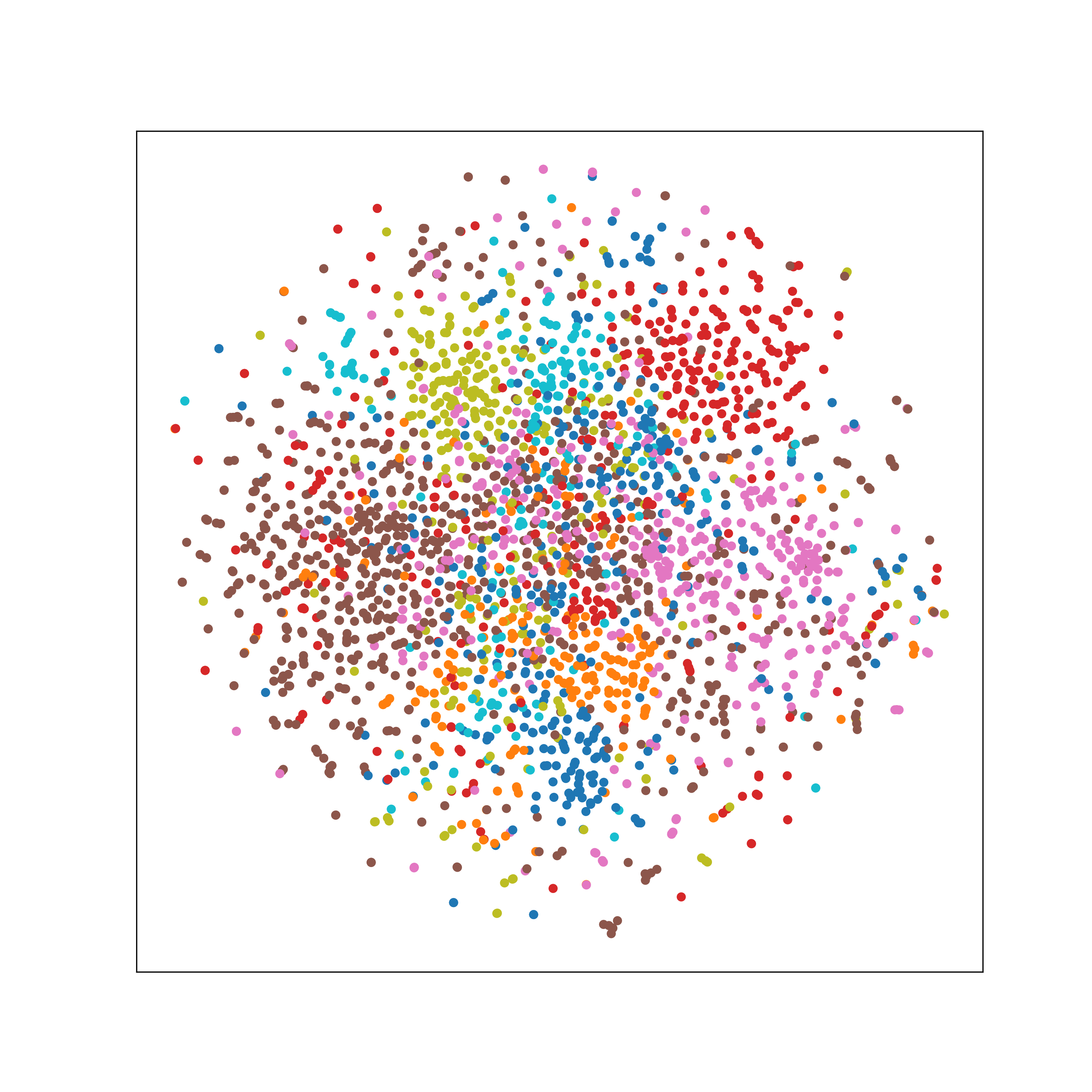}
\label{g2_1}
}
\subfigure[gCooL (ACC=72.3)]{
\includegraphics[width=0.31\textwidth]{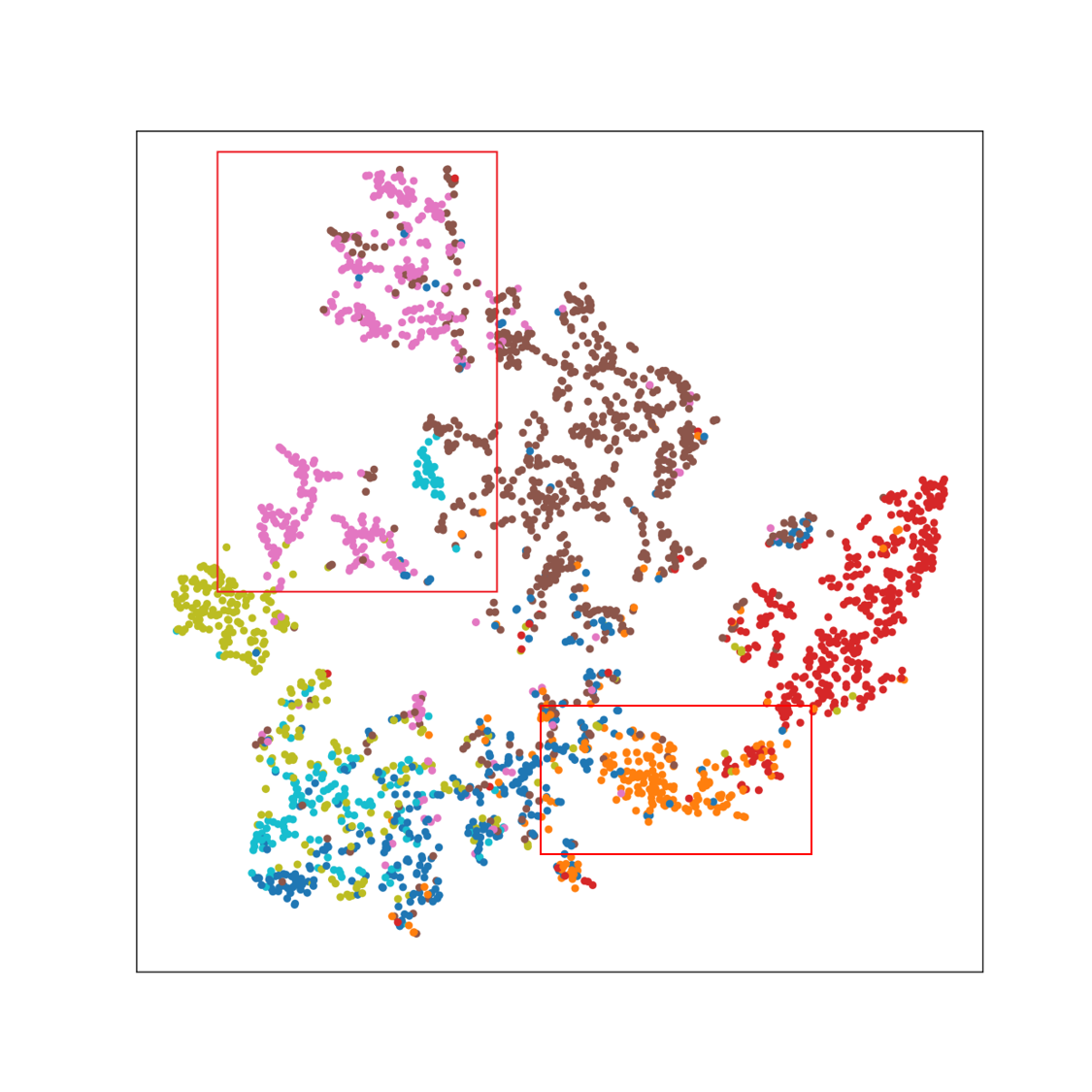}
\label{g2_2}
}
\subfigure[Ours (ACC=77.9)]{
\includegraphics[width=0.31\textwidth]{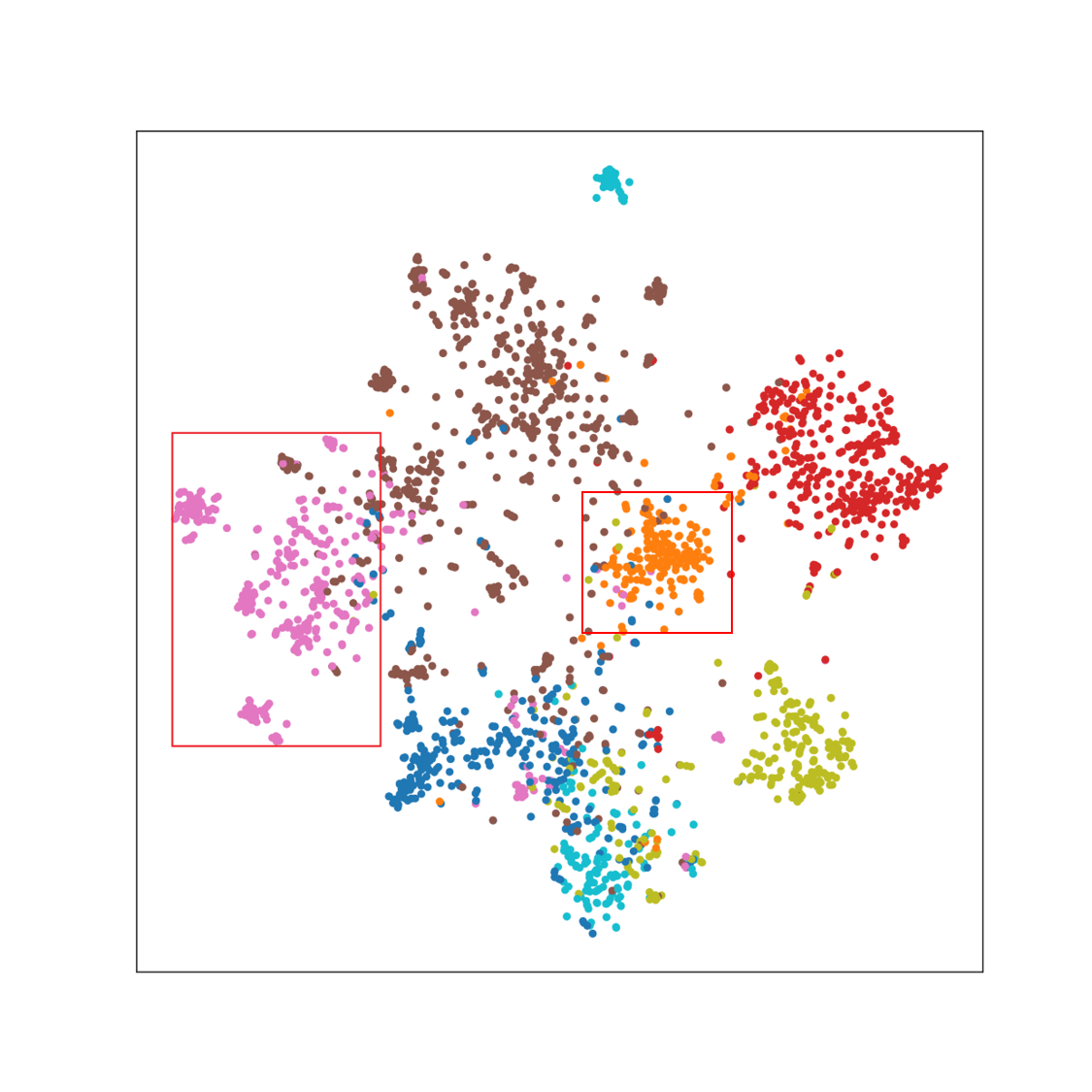}
\label{g2_3}
}
\caption{The t-SNE visualizations on Cora. (a) is the visualization for raw node features. (b) is the visualization for the node representations from gCooL. (c) is the visualization for the node representations from our CueGCL. The node representations generated by our CueGCL are more distinguishable than the gCooL.}
\label{fg2}
\end{figure*}

 Recently, graph representation learning has become an important backbone for graph clustering that incorporates network topology and node attributes \cite{xiao2024GraphACL}. Especially, graph contrastive learning (GCL) is a superior technique to enhance the discriminability of nodes in graph representation learning by comparing positive and negative samples. It has demonstrated remarkable success in semi-supervised downstream tasks such as node classification \cite{li2022gCool}, graph classification \cite{li2024IJCAI}, and link prediction \cite{zhang2022line}. Nonetheless, within the context of graph representation learning utilizing GCL, two significant challenges persist:

 \begin{tcolorbox}
 \textbf{Challenge \MakeUppercase{\romannumeral 1} \quad Cluster-Level Awareness of Graphs} 
 \end{tcolorbox}
 Some structure-related and unsupervised learning tasks in network analysis also attract much attention, such as graph clustering. \textbf{Most graph contrastive learning frameworks have difficulty learning the structural information required in these problems.} They focus only on node-level similarity but fail to consider the cluster structure in the graph when sampling positive and negative node pairs.
 To further illustrate the importance of cluster-level information, we use t-SNE to visualize node representations from a novel GCL algorithm \emph{gCooL} \cite{li2022gCool} and our \emph{CueGCL} in Figure \ref{fg2}, and report their accuracy (ACC) of graph clustering. gCooL is a Graph Communal Contrastive Learning algorithm that can learn both node representations and community partition. Compared with Figures \ref{g2_2} and \ref{g2_3}, we observe that the node representation space of gCooL does not have good community properties, and also its clustering accuracy is significantly lower than ours. Specifically, our model pays attention to learning the personalized features of each community to obtain a compact embedding space within the clusters, also as shown in Figure \ref{g6}, but existing methods fail to do so.

\begin{table}[!ht]
\renewcommand{\arraystretch}{1.5}
\centering
\begin{center}
\scalebox{0.9}{
\begin{tabular}{|@{\extracolsep{1pt}} c | c c c c c|}
\hline
$N_{neg}$ & 50 & 100 & 500 & 1000 & 1500 \\ [0.5ex]
\hline\hline
Micro-F1 & 32.3 & 23.4 & 18.1 & 6.9 & 4.6\\
\hline
Macro-F1 & 31.3 & 25.5 & 19.3 & 5.4 & 3.9\\
\hline
Number of nodes in $4^{th}clu$ & 380 & 343 & 54 & 34 & 5\\
\hline
time / epoch (s) & 0.6 & 0.7 & 0.9 & 1.0 & 1.2\\ 
\hline
\end{tabular}
}
\caption{Effect of negative sample size on contrastive-based models in graph clustering task on Citeseer.}
\label{t_1}
\end{center}
\end{table}

\begin{tcolorbox}
\textbf{Challenge \MakeUppercase{\romannumeral 2} \quad Fair Awareness for Each Cluster} 
\end{tcolorbox}
Some contrastive-based methods typically require a large number of negative samples, allowing for uniform separation of representations across all samples. However, the contrast of large negative samples would inevitably lead to the class collision issue, which means that different samples from the same class are treated as negative sample pairs and pushed away incorrectly \cite{saunshi2019theoretical}. Moreover, we find that \textbf{class collision induces unfairness in unsupervised graph clustering, which lacks label information to mitigate bias in negative sampling of GCL compared to supervised downstream tasks.} Specifically, certain clusters may not be aware of the model, leading to inaccurate detection. To further explain, we conduct graph clustering with the general GCL model on Citeseer and report the experimental results in Table \ref{t_1}. We empirically observe that as the negative sample size (i.e. denoted as $N_{neg}$) increases, the F1 score shows a substantial decrease. Furthermore, the number of nodes that the model marks as the fourth cluster (i.e. $4^{th}clu$), named 'ML', on the Citeseer keeps decreasing. This means that almost all nodes in the 'ML' cluster (actually should be 596) are pulled into other clusters by the contrastive-based model because they are incorrectly identified as negative sample pairs. A good model should fairly perceive each cluster in the graph, meaning that the accuracy of any single cluster should not be significantly lower than that of other clusters or the overall clustering accuracy of the graph. Overall, we conclude that the contrastive-based model cannot fairly perceive each cluster in the network.

To address the above challenges, we propose a novel GCL method, called \textbf{\underline{Com}}munity-awar\textbf{\underline{e}} \textbf{\underline{G}}raph \textbf{\underline{C}}ontrastive \textbf{\underline{L}}earning (\textbf{CueGCL}).

The main contributions of our paper are summarized as follows:
\begin{itemize}
\item We present an end-to-end graph contrastive learning framework, called CueGCL, that jointly trains cluster partition and node embeddings.
\item We propose a personalized self-training (PeST) strategy for unsupervised scenarios that allows our model to actively learn the personalized features of each cluster. Moreover, it is theoretically demonstrated that our PeST strategy adaptively results in a more compact internal structure of each cluster.
\item We point out that our PeST module is actually more unbiased for negative sampling, thereby effectively alleviating the class collision problem. Furthermore, by incorporating balanced sampling in PeST, our model enhances its fair aware ability for each cluster.
\item Experimental results over the five benchmark datasets indicate that our CueGCL outperforms the state-of-the-art. Moreover, the effectiveness of our model for mitigating the unfairness problem is verified.
\end{itemize}

\begin{figure*}[t]
\centering
\includegraphics[width=6.1in]{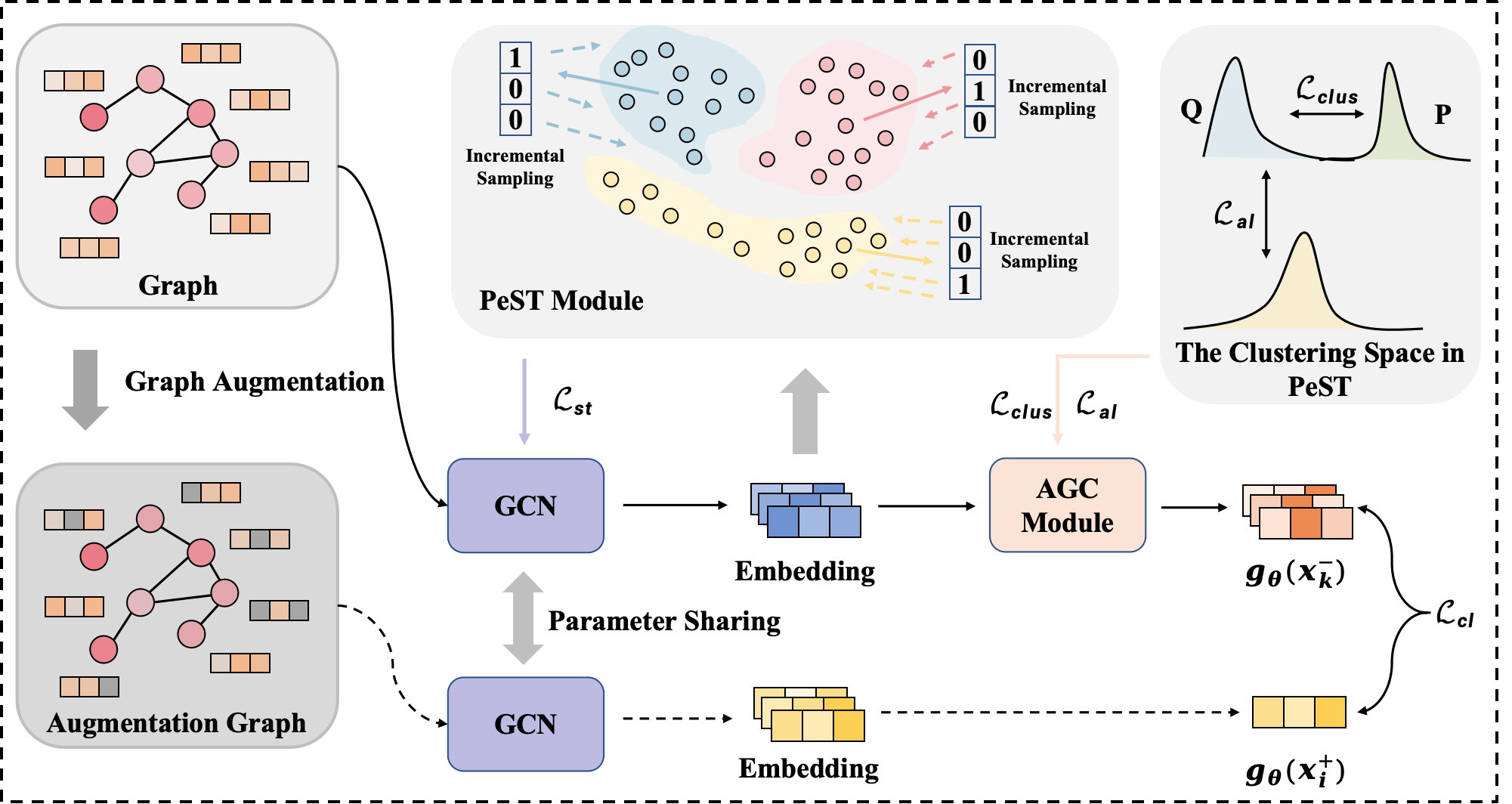}
\caption{The overall framework of our CueGCL. It consists of three parts: graph contrastive learning framework, personalized self-training (PeST) module, and aligned graph clustering (AGC) module. In our framework, we use the feature-space graph augmentation strategy, where $x_i^{+}$ represents the positive sample and $x_{k}^{-}$ represents the negative sample if $k \in \widetilde{\mathcal{N}}_i$. The clustering results and node embeddings are both optimized by objective functions: $\mathcal{L}_{cl}, \mathcal{L}_{st}, \mathcal{L}_{clus}$ and $\mathcal{L}_{al}$.}
\label{g3}
\end{figure*}

\section{Related Work}

\textbf{Graph Contrastive Learning.} Graph contrastive learning is a powerful self-supervised learning technique that focuses on learning representations of nodes in the semantic space by comparing positive and negative samples. Contrasts between positive samples enable representations to learn consistency across different views of similar samples, while contrasts with negative samples emphasize the differences between different classes of samples. DGI \cite{velickovic2019DGI} uses an objective function based on mutual information rather than random walking, which applies the ideas of DIM \cite{hjelm2018learning} to graph data so that the node representations contain global graph information. MVGRL \cite{hassani2020MVGRL} introduces different structural views of a graph based on DGI for learning both node-level and graph-level representations. GRACE \cite{zhu2020GRACE} proposes a simplified contrastive framework by using InfoNCE loss\cite{oord2018infoNCE} with both inter-view and intra-view negative pairs. BGRL \cite{thakoor2021BGRL} introduces a graph self-supervised learning method that uses two encoders to generate node representations without negative samples. However, most non-contrastive methods still depend on complex architectures or dataset-specific augmentations \cite{DBLP:conf/kdd/ShiaoSL0SP23}. Overall, above mentioned contrastive-based methods are not suitable for graph clustering directly as they only focus on node-level or graph-level representation learning and ignore the specific cluster structure of the network.

\noindent \textbf{Deep Graph Clustering. }
More recently, graph representation learning combined with clustering algorithms has become a popular trend in graph clustering. MinCutPool \cite{bianchi2020MinCutPool} proposes a continuous relaxation graph pooling method that leverages GNNs to learn node clustering, avoiding the computational cost of spectral decomposition. DMoN \cite{tsitsulin2023DMoN} presents a novel GNN pooling method that optimizes modularity for efficient unsupervised graph clustering. ARVGA \cite{pan2018arga} uses a graph autoencoder to map the node on the attribute graph into a compact representation, and combines an adversarial training module to make it conform to a prior distribution. VGAER \cite{qiu2022VGAER} proposes a variational graph autoencoder for reconstruction-based graph clustering, and gives its non-probabilistic version. However, such reconstruction-based algorithms extract features at a fine granularity, so they are unsuitable for downstream node-level tasks, e.g., node classification and clustering. CommDGI \cite{zhang2020commdgi} introduces the trainable clustering layer and community-oriented objectives, which learns the community-related node representations through joint optimization. GDCL \cite{zhao2021GDCL} combines a clustering layer with the contrastive learning framework, which utilizes a debiased negative sampling strategy to reduce the false negative samples. gCooL \cite{li2022gCool} proposes the reweighted self-supervised cross-contrastive training module to reduce the bias in contrastive learning using community information. $S^{3}$-CL \cite{ding2023S3CL} infers node clusters and prototypes, and enforces nodes in the same cluster to group around their corresponding prototypes in the latent space. However, graph prototypical contrastive learning does not account for the sampling bias in positives. Moreover, strictly enforcing intra-cluster compactness may reduce the continuity and generalization capability of the representation space. In general, the existing graph representation algorithms based on contrastive learning do not solve the class collision and unfairness problem well, especially when the number of negative samples is large, which would hamper the effect of graph clustering.


\section{Proposed Method}
In this section, we detail how our proposed CueGCL achieves graph clustering. Specifically, we first construct a basic \emph{Graph Contrastive Learning} framework in Section \ref{subsec: EGCL}. Next, based on the node embeddings encoded by GCN in Graph Contrastive Learning framework, we design a \emph{Personalized Self-Training} strategy in Section \ref{subsec:ST}. Meanwhile, we introduce a \emph{clustering layer} on the embeddings from GCN to achieve the clustering of the downstream task (Section \ref{CL}), and \emph{align} the clustering space of the downstream task with the clustering space of the PeST (Section \ref{AM}). 
The general framework of our model is shown in Figure \ref{g3}. 

Firstly, we introduce some notations used in this paper. Given a network $\mathcal{G=(V,E,X)}$, where $\mathcal{V}=\{v_1,v_2,\ldots,v_n\}$ denotes the set of nodes in network, $\mathcal{E}=\{e_{ij}\}\subseteq \mathcal{V} \times \mathcal{V}$ denotes the set of edges between nodes, $\mathcal{X}=\{\boldsymbol{x_1}, \boldsymbol{x_2},\ldots,\boldsymbol{x_n}\}$ denotes the set of node attributes and $\boldsymbol{x_i} \in \mathbb{R}^d (i=1,2, \ldots, n)$ is the feature of $v_i$, $d$ is the dimension of the node attribute. We denote the adjacency matrix of the network as \( \boldsymbol{A} \in \mathbb{R}^{n \times n} \) and the feature matrix as \( \boldsymbol{X} \in \mathbb{R}^{n \times d} \).

\subsection{Graph Contrastive Learning}
\label{subsec: EGCL}
Graph contrastive learning obtains high-quality semantic representations by pulling positive sample pairs $(\boldsymbol{x_i},\boldsymbol{x_i^{+}})$ closer and negative sample pairs $(\boldsymbol{x_i},\boldsymbol{x_k^{-}})$ further away in the embedding space.  Thus, CueGCL minimizes the following objective function for every node in the graph:


\begin{equation}
\begin{aligned}
    & \mathcal{L}_{c l}\left(\boldsymbol{x}_i\right) =-\ln \frac{e^{{g_{\theta}\left(\boldsymbol{x_i}\right)^T g_{\theta}\left(\boldsymbol{x_i^{+}}\right)} / {\tau}}}{e^{{g_{\theta}\left(\boldsymbol{x_i}\right)^T g_{\theta}\left(\boldsymbol{x_i^{+}}\right)} / {\tau}}+\sum\limits_{k \sim \widetilde{\mathcal{N}}_i} e^{{g_{\theta}\left(\boldsymbol{x_i}\right)^T g_{\theta}\left(\boldsymbol{x}_{k}^{-}\right)} / {\tau}}},
\end{aligned}
\end{equation}
where $g_{\boldsymbol{\theta}}(\cdot)$ is a GCN \cite{kipf2016GCN} with its parameter $\boldsymbol{\theta}$ and $\tau$ is temperature parameter. We mask node feature $\boldsymbol{x}_i$ randomly and obtain the positive sample $\boldsymbol{x}_i^{+}$, which is a kind of graph augmentation strategy. Here, $\widetilde{\mathcal{N}}_i = \{v_k\}(k \neq i)$ is the negative sample set of node $v_i$. 

\subsection{Personalized Self-Training}
\label{subsec:ST}
\begin{figure}[!t]
    \centering
    \includegraphics[width=2.8in]{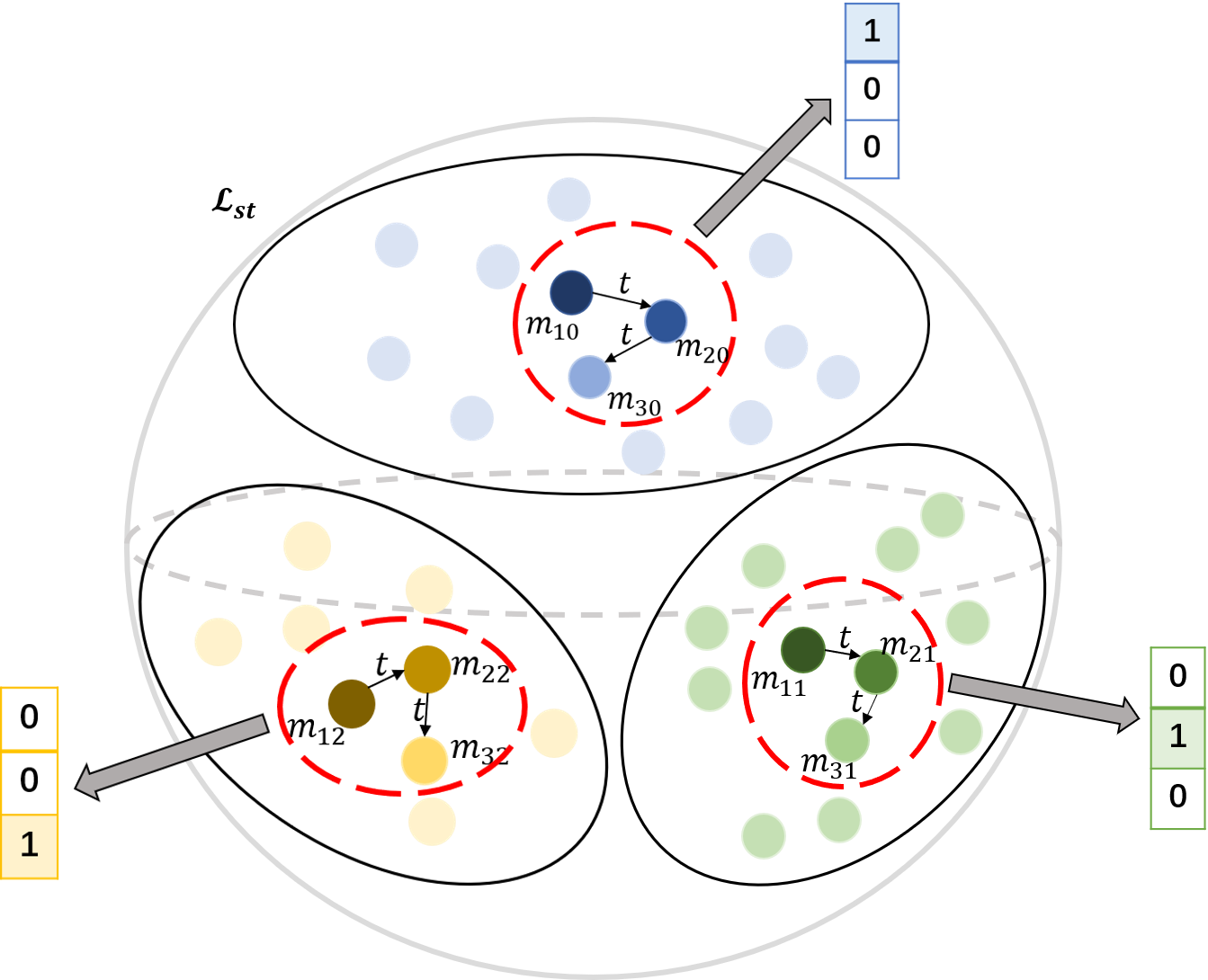}
    \caption{An illustration of incremental sampling and personalized self-training strategy. The deeper the color of the medoid, the more central it is.}
    \label{g4}
\end{figure}

Self-training intends to extract semantic information from unlabeled data by adding high-confidence prediction nodes and their labels from each class to the training set. Inspired by it, we propose an incremental sampling and personalized self-training (PeST) strategy in unsupervised scenarios, as shown in Figure \ref{g4}. Moreover, we demonstrate its effectiveness in experimental and theoretical aspects, respectively.

\noindent \textbf{Implementation details.}
In contrast to general self-training algorithms that explore high-confidence node prediction results \cite{sun2020M3S}, we choose to sample the medoids of each cluster in the embedding space for self-training and add them to the training set incrementally. Particularly, after each $t$ epoch (i.e. the sampling frequency of medoids), one sampling is performed in the embedding space, and the obtained $K$ medoid vectors are added to the training set, where $K$ refers to the number of clusters in the graph and can be obtained by some non-parametric methods \cite{kulis2011un_kmeans,jiang2011un_clustering}. We use the K-medoids \cite{park2009K-medoids} algorithm to obtain the real node $v_j$ in each cluster center with the following objective function:
\begin{equation}\label{eq2}
\sum_{i=1}^n \min _{j \in U_l} d\left(v_i, v_j\right)=\sum_{i=1}^n \min _{j \in U_l}\left\|g_{\theta}\left(\boldsymbol{x}_i\right)-g_{\theta}\left(\boldsymbol{x}_j\right)\right\|_2,
\end{equation}
where $U_l=U_{l-1} \setminus M_{l-1}$ is the set of nodes that are not selected as the medoid in $(l*t)$-th epoch and $t$ represents the sampling frequency of medoids, $l$ represents the sampling times. Also, we obtain $M_l=\{m_{l,0},m_{l,1},\ldots,m_{l,K-1}\}$ as the selected $K$ medoids in $(l*t)$-th epoch. In graph clustering scenario, we do not have any label information in advance. Thus the subscript corresponding to the medoid vector is directly assigned to its own label (i.e. $label(m_{l, i})=i$), and the label set is denoted as $N_l=\{0,1,\ldots,K-1\}$. Therefore, in the $T$-th (i.e. $T=L*t$) epoch, we obtain the training set $M^T=\bigcup_{l=1}^L{M_l}$ and its label set $N^T=\bigcup_{l=1}^L{N_l}$. In order to learn the distribution of personalized features for each cluster, our CueGCL minimizes the following cross-entropy objective function $\mathcal{L}_{st}$:
\begin{equation} 
    \mathcal{L}_{st}=-\frac{1}{|M^T|}\sum_{i \in M^T}\sum_{n=0}^{K-1}y_{i, n}\ln{\hat{y}_{i, n}}.  \label{eq:Lst}
\end{equation}
Here, $y_{i,n}=1$ if node $i$ is the medoid of cluster $n$ and 0 otherwise, indicating the $n$-th element of the one-hot label vector $\boldsymbol{y}_i$. $\hat{y}_{i, n}$ is the $n$-th element of the prediction label vector $\boldsymbol{\hat{y}}_i$ and it calculates as:
\begin{equation}  \boldsymbol{\hat{y}}_i=MLP(g_\theta(\boldsymbol{x}_i)),
\end{equation}
where $MLP$ is a two-layer fully connected neural network. Here, the role of the $MLP$ is to generate the predicted label with semantic information. In addition, the incremental sampling strategy allows the more central medoids to be added in the training set earlier, thus our model could capture the personalized features of each cluster more accurately, as shown in Figure \ref{g4}.

The reason why we choose to sample the medoids of each cluster for self-training is that this approach satisfies the following three essential principles:
\begin{enumerate}
    \item The set of nodes is selected to be representative. General self-training algorithms tend to choose high-confidence nodes to join the training set, yet these nodes do not necessarily contain the critical information to train, especially the cluster-level semantic information. K-medoids can obtain better cluster centers, as theoretically demonstrated in \cite{wu2019feapro}, thus allowing our model to capture cluster-level personalized information effectively.
    \item The class information of the selected nodes is relatively precise. We select the medoids of each cluster since the confidence of their class is higher than other nodes. From Figure \ref{g4}, we could show that our PeST module can be regarded as the contrastive learning of debiased negative samples, considering the selection of the medoids with less bias. It achieves a more accurate cluster-level discriminative role compared to the general contrastive learning algorithm. Thus our model only needs a small number of negative samples in GCL.
    \item The set of nodes is selected to ensure class balance. The balanced sampling of medoids allows our model to learn more equally the information of each cluster in the graph, which guarantees the fairness of graph clustering to a degree.
\end{enumerate}

More details about the PeST module, including its effectiveness and scalability as well as the reasons for using K-medoids, can be found in the \textbf{Appendix \ref{app:G}} \footnote{For detailed information on the appendix, please refer to the uploaded \textbf{supplementary materials}.}.

\noindent \textbf{Experimental and Theoretical Analysis.}
In the following, we will demonstrate experimentally and theoretically the effectiveness of our self-training strategy.

\begin{figure}[t]
    \centering
    \includegraphics[width=2.6in]{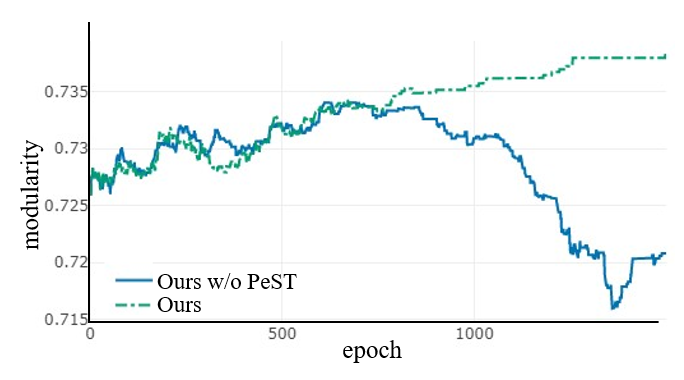}
    \caption{Effectiveness of our PeST module on Citeseer.}
    \label{g6}
\end{figure}

In Figure \ref{g6}, we study the effectiveness of our PeST module by \emph{modularity} metric, which measures the compactness of the cluster structure \cite{newman2006modularity}. It can be observed that with the removal of the self-training module from our CueGCL, the modularity tends to decrease as training, which validates that our PeST module allows for a more compact structure within the clusters. More specifically, our model learns the personalized features of each cluster, so its embedding space has better cluster properties than those of the general contrastive learning method. 

In theory, our PeST strategy is equivalent to adaptively drawing closer nodes of the same cluster, as formally proposed in Theorem \ref{Thm3.1}.

\begin{tcolorbox}
\begin{theorem}
\label{Thm3.1}
An undirected graph $\mathcal{G}$ has $n$ nodes $\mathcal{V}=\left\{v_1, \ldots, v_n\right\}$, and $\boldsymbol{H}_i^{(t)}$ $\left(\boldsymbol{H}_i^{(0)}=\boldsymbol{x}_i\right)$ is the embedding of node $v_i$ after $t$ times message propagating and aggregation. And $\boldsymbol{y}_k \in \mathbb{R}^K$ is an one-hot label vector, where $y^{(k)}=1$ is the $k$-th element of $\boldsymbol{y}_k$. Assume that node $v_i$ belongs to cluster $k$, then we have $\forall \epsilon >0, \exists T \in \mathbb{N}^{+}$, such that $\forall t>T$, then $\left\|\boldsymbol{H}_i^{(t)}-\boldsymbol{y}_k\right\|_2 \leq \epsilon$.
\end{theorem}
\end{tcolorbox}
\begin{proof}
Please refer to \textbf{Appendix \ref{A1}} for the complete proof.
\end{proof}
 The above theorem demonstrates that all nodes of the same cluster adaptively converge to a specified center. These center vectors are two-by-two orthogonal and span to a standard orthogonal space. 

\textbf{Summary:} We use the K-medoids algorithm to incrementally extract the medoids of each cluster such that the clustering space converges to a set of standard orthogonal bases in the Euclidean space by optimizing the cross-entropy loss. The PeST module allows us to obtain a node embedding space that is \textbf{tight within the cluster} and \textbf{sparse between the clusters}.

\subsection{Aligned Graph Clustering}
\label{subsec:GC}
In this subsection, we propose an aligned graph clustering layer and jointly optimize it with graph contrastive learning and personalized self-training module. Based on the DEC algorithm \cite{xie2016DEC}, we add the learnable cluster centers in downstream task to the self-training, in order to align them with the cluster medoids in PeST. 

\noindent \textbf{Clustering Layer.}
\label{CL}
To obtain the graph clustering results, we develop a clustering layer. At first, we compute a soft assignment between the node embedding $g_\theta(\boldsymbol{x}_i)$ and the cluster center parameter $\boldsymbol{\mu}_k$. The similarity between them is measured by the Student’s $t$-distribution $Q$ as follows:
\begin{equation} \label{eq5}
q_{ik}=\frac{\left(1+\left\|g_{\boldsymbol{\theta}}\left(\boldsymbol{x}_i\right)-\boldsymbol{\mu}_k\right\|^2 / \alpha \right)^{-\frac{\alpha + 1}{2}}}{\sum_j\left(1+\left\|g_{\boldsymbol{\theta}}\left(\boldsymbol{x}_i\right)-\boldsymbol{\mu}_k\right\|^2 / \alpha \right)^{-\frac{\alpha + 1}{2}}},
\end{equation}
where $\alpha$ is the degree of freedom of the t-distribution and $q_{ik}$ can be interpreted as the probability of assigning node $i$ to cluster $k$. We normally set $\alpha=1$ and initialize the cluster centers $\{\boldsymbol{\mu}_k\}_{k=1}^{K}$ by K-means. Secondly, we define the $L_{clus}$ as the KL divergence loss between the soft assignments $Q$ and the target distribution $P$ as follows:
\begin{equation}
L_{clus}=K L(P \| Q)=\sum_i \sum_k p_{i k} \log \frac{p_{i k}}{q_{i k}},
\end{equation}   
where $p_{i k}$ is the target distribution $P$ with higher confidence assignments defined as follows:
\begin{equation} \label{eq7}
p_{i k}=\frac{q_{i k}^2 / f_k}{\sum_{k} q_{i k}^2 / f_{k }},
\end{equation}
where $f_k = \sum_i q_{ik}$ is the soft cluster frequency.

\begin{table*}[!t]
\renewcommand{\arraystretch}{1.1}
\centering
\begin{center}
\resizebox{2.0\columnwidth}{!}{
\begin{tabular}{clccccccccccccc}
\hline {} & \multirow{2}{*}{ Method } & \multirow{2}{*}{ Input } & \multicolumn{2}{c}{ Cora } & \multicolumn{2}{c}{ Citeseer } & \multicolumn{2}{c}{ PubMed }  & \multicolumn{2}{c}{ Cornell }  & \multicolumn{2}{c}{ Texas } & \multicolumn{2}{c}{ \textit{Avg. Rank} } \\
\cline { 4-15 } & {} &{} & ACC & NMI & ACC & NMI & ACC & NMI & ACC & NMI & ACC & NMI & ACC & NMI \\
\hline 
\multirow{2}{*}{ \textbf{TA} } & SC & $\mathbf{A}$ & $36.7$ & $12.6$ & $23.8$ & $5.5$ & $52.8$ & $9.7$ & $23.6$ & $2.6$ & $22.1$ & $6.4$ & $16.2$ & $16.8$ \\
{} & $K$-means & $\mathbf{X}$ & $49.2$ & $32.1$ & $54.0$ & $30.5$ & $59.5$ & $31.5$ & $21.3$ & $3.6$ & $23.4$ & $10.4$ & $14.4$ & $11.6$ \\
\hline 
\multirow{2}{*}{ \textbf{GNN-A} } & MinCutPool (ICML'20) & $(\mathbf{A}, \mathbf{X})$ & $49.0$ & $41.0$ & $53.7$ & $29.5$ & $52.1$ & $21.4$ & $42.1$ & $11.9$ & $50.5$ & $11.8$ & $10.8$ & $10.2$ \\
{} & DMoNn (JMLR'23) & $(\mathbf{A}, \mathbf{X})$ & $51.7$ & $47.3$ & $38.5$ & $30.3$ & $35.1$ & $25.7$ & $42.6$ & $\underline{12.5}$ & $46.3$ & $9.9$ & $11.0$ & $9.4$ \\
\hline 
\multirow{6}{*}{ \textbf{GA} } & VGAE (NIPS'16) & $(\mathbf{A}, \mathbf{X})$ & $50.2$ & $32.9$ & $46.7$ & $26.0$ & $63.0$ & $22.9$ & $39.7$ & $6.9$ & $38.5$ & $13.5$ & $12.2$ & $12.0$ \\
{}& ARGA (IJCAI'18) & $(\mathbf{A}, \mathbf{X})$ & $64.0$ & $44.9$ & $57.3$ & $35.0$ & $66.8$ & $30.5$ & $36.4$ & $7.1$ & $42.5$ & $15.5$ & $11.2$ & $9.0$ \\
{} & ARVGA (IJCAI'18) & $(\mathbf{A}, \mathbf{X})$ & $64.0$ & $45.0$ & $54.4$ & $26.1$ & $69.0$ & $29.0$ & $38.6$ & $8.3$ & $41.3$ & $\underline{15.8}$ & $9.6$ & $8.8$ \\
{} & DAEGC (IJCAI'19) & $(\mathbf{A}, \mathbf{X})$ & $69.4$ & $50.6$ & $64.5$ & $36.4$ & $68.7$ & $28.3$ & $35.0$ & $7.4$ & $52.3$ & $12.4$ & $7.4$ & $8.8$ \\
{} & VGAER (AAAI'22) & $(\mathbf{A}, \mathbf{X})$ & $45.3$ & $29.7$ & $30.2$ & $21.7$ & $30.1$ & $22.3$ & $29.5$ & $6.2$ & $26.2$ & $3.0$ & $15.8$ & $15.6$ \\
{} & DDGAE (ESWA'24) & $(\mathbf{A}, \mathbf{X})$ & $73.7$ & $55.5$ & $68.2$ & $41.7$ & $69.1$ & $\underline{33.1}$ & $32.7$ & $6.1$ & $44.5$ & $11.3$ & $7.0$ & $7.6$ \\
\hline 
\multirow{6}{*}{ \textbf{CA} } & CommDGI (CIKM'20) & $(\mathbf{A}, \mathbf{X})$ & $66.1$ & $52.8$ & $62.1$ & $39.8$ & $63.2$ & $30.6$ & $39.3$ & $7.2$ & $53.0$ & $11.4$ & $7.8$ & $8.2$ \\
{} & MVGRL (ICML'20) & $(\mathbf{A}, \mathbf{X})$ & $73.6$ & $58.8$ & $68.1$ & $43.2$ & $\underline{69.2}$ & $32.0$ & $38.3$ & $8.9$ & $41.5$ & $9.5$ & $7.2$ & $\underline{6.0}$ \\
{} & GDCL (IJCAI'21) & $(\mathbf{A}, \mathbf{X})$ & ${74.7}$ & ${58.9}$ & $\underline{69.7}$ & ${44.8}$ & $68.4$ & ${32.7}$ & $\underline{45.3}$ & $7.9$ & $\underline{55.7}$ & $8.5$ & $\underline{3.4}$ & $\underline{6.0}$ \\
{} & gCooL (WWW'22) & $(\mathbf{A}, \mathbf{X})$ & $72.3$ & $55.8$ & $63.8$ & $34.6$ & $67.8$ & $30.3$ & $33.9$ & $7.6$ & $43.7$ & $9.3$ & $9.2$ & $9.4$ \\
{} & $S^{3}$-CL (AAAI'23) & $(\mathbf{A}, \mathbf{X})$ & $73.9$ & $58.0$ & $68.8$ & $43.7$ & $69.0$ & $32.3$ & $42.7$ & ${9.4}$ & $50.3$ & $8.1$ & $4.4$ & $6.2$ \\
{} & HSAN (AAAI'23) & $(\mathbf{A}, \mathbf{X})$ & $\underline{75.1}$ & $\underline{59.0}$ & $69.2$ & $\underline{45.9}$ & $68.6$ & $31.5$ & $41.2$ & $8.1$ & $52.8$ & $7.6$ & $4.4$ & $6.4$ \\
\hline 
\textbf{CA} & CueGCL (Ours) & $(\mathbf{A}, \mathbf{X})$ & $\mathbf{77.9}$ & $\mathbf{60.5}$ & $\mathbf{71.4}$ & $\mathbf{47.2}$ & $\mathbf{71.2}$ & $\mathbf{36.8}$ & $\mathbf{46.1}$ & $\mathbf{13.6}$ & $\mathbf{56.7}$ & $\mathbf{17.1}$ & $\mathbf{1.0}$ & $\mathbf{1.0}$ \\
{} & Gain & & $\mathbf{+2.8}$ & $\mathbf{+1.5}$ & $\mathbf{+1.7}$ & $\mathbf{+1.3}$ & $\mathbf{+2.0}$ & $\mathbf{+3.7}$ & $\mathbf{+0.8}$ & $\mathbf{+1.1}$ & $\mathbf{+1.0}$ & $\mathbf{+1.3}$ & $\mathbf{+2.4}$ & $\mathbf{+5.0}$\\
\hline
\end{tabular}
}
\caption{The average performance of graph clustering with ACC and NMI on Cora, Citeseer, PubMed, Cornell, and Texas. The bold and \underline{underlined} text indicates the optimal and suboptimal results, respectively. The \textit{Avg. Rank} denotes the mean value across all five datasets for the rank among all methods about a specific metric.}
\label{t_2}
\end{center}
\end{table*}

\noindent \textbf{Alignment Module.}
\label{AM}
The clustering space of the downstream task is inconsistent with that in PeST, which will cause the information in PeST to be useless or even harmful to our downstream task. So we align these two spaces to reduce the interference of this inconsistent information on downstream tasks. Specifically, we add $C = \{\boldsymbol{\mu}_k\}_{k=1}^{K}$ to the training set for self-training, similar to Section \ref{subsec:ST}. Then, we minimizes the following cross-entropy objective function $\mathcal{L}_{al}$:
\begin{equation} \label{Lal}
    \mathcal{L}_{al}=-\frac{1}{|C|}\sum_{i \in C}\sum_{n=0}^{K-1}y_{i,n}\ln{\hat{y}_{i,n}}.  
\end{equation}
Here, $y_{i,n}=1$ if $i = n$ and 0 otherwise, indicating the $n$-th element of the one-hot label vector $\boldsymbol{y}_i$. $\hat{y}_{i,n}$ is the $n$-th element of the prediction label vector $\boldsymbol{\hat{y}}_i$ and it calculates as:
\begin{equation}
    \boldsymbol{\hat{y}}_i=MLP(\boldsymbol{\mu }_i).
\end{equation}
We align the clustering space in the downstream task with that in the representation space, which achieves more consistent node embeddings. Moreover, the cluster centers $\{\boldsymbol{\mu}_k\}_{k=1}^{K}$ are learnable, which means they can be optimized continuously during the training process. This provides additional training paradigms for our CueGCL to learn the personalized characteristics of each cluster, thereby enhancing the generalization capability of our model.


\subsection{Joint Optimization}
Consequently, we jointly graph contrastive learning, personalized self-training, and aligned graph clustering in an end-to-end trainable framework for more salient cluster structures. The overall objective function of the proposed CueGCL is induced as:
\begin{equation} \label{L}
    \min \mathcal{L} = \frac{1}{n}\sum_{i=1}^n\mathcal{L}_{cl}\left(\boldsymbol{x}_i\right) + \gamma_{st}\mathcal{L}_{st} + \gamma_{clus}\mathcal{L}_{clus} + \gamma_{al}\mathcal{L}_{al},  
\end{equation}
where $\gamma_{st}$, $\gamma_{clus}$, $\gamma_{al} > 0$ are trade-off parameters to balance the magnitudes between different loss items. 

On the one hand, the node representations obtained by contrastive learning have better discriminative properties and improve the accuracy of class information, thus facilitating the PeST module. On the other hand, PeST makes the contrastive learning of negative samples more accurate and efficient. With the mutual promotion of the various modules in our CueGCL, we can both obtain discriminative node representations and fair graph clustering. The whole algorithm is summarized in Algorithm 1 in \textbf{Appendix \ref{A5}}.


\section{Experiment}

\subsection{Experimental Setup}
\textbf{Datasets. }In this paper, we perform experiments on various real networks with different scales widely used in unsupervised graph representation learning \cite{zhao2021GDCL,yuan2023muse}. The five benchmark datasets are: Cora \cite{sen2008cora}, Citeseer \cite{rossi2015citeseer}, PubMed \cite{namata2012pubmed}, Cornell and Texas \cite{pei2020cornell}. Their detailed information is shown in \textbf{Appendix \ref{A2}}.

\noindent \textbf{Baseline. }In order to verify the superiority of our model, we compare it with the current state-of-the-art algorithms as follows: (1) \textbf{Unsupervised Algorithms}, including (1.1) Traditional algorithms (\textbf{TA}): Spectral clustering (SC) \cite{amini2013SPECTRAL_CLUSTERING}, K-means \cite{K-MEANS}; (1.2) GNN-based algorithms (\textbf{GNN-A}): MinCutPool \cite{bianchi2020MinCutPool}, DMoN \cite{tsitsulin2023DMoN}; (1.3) Generative-based algorithms (\textbf{GA}): VGAE \cite{kipf2016VGAE}, ARGA \cite{pan2018arga}, ARVGA \cite{pan2018arga}, VGAER \cite{qiu2022VGAER}, DAEGC \cite{wang2019DAEGC}, DDGAE \cite{wu2024DDGAE}; (1.4) Contrastive-based algorithms (\textbf{CA}): CommDGI \cite{zhang2020commdgi}, MVGRL \cite{hassani2020MVGRL}, GDCL \cite{zhao2021GDCL}, gCooL \cite{li2022gCool}, $S^{3}$-CL \cite{ding2023S3CL}, HSAN \cite{liu2023HSAN}. (2) \textbf{Semi-Supervised Algorithms}, including Self-training algorithms (\textbf{SA}): Co-training \cite{li2018self-training}, Union \cite{li2018self-training}, M3S \cite{sun2020M3S}.


\noindent \textbf{Evaluation Metrics.} Two evaluation metrics are used to measure the performance of various algorithms, namely Accuracy (ACC) and Normalized Mutual Information (NMI). In addition, we adopt the F1 score (Micro-F1 and Macro-F1) to evaluate the fairness of clustering, which can comprehensively reflect the detection accuracy for each cluster.

\noindent \textbf{Parameters Settings. }Our experiments are run on the machine with two NVIDIA GeForce RTX 3090 Ti and the details of the hyperparameters are provided in \textbf{Appendix \ref{A3}}.

\subsection{Graph Clustering}
We conduct the graph clustering experiment with our CueGCL and various comparison methods on five datasets 20 times and report the average results. We show the experimental results in Table \ref{t_2}, \ref{t_4}, and \ref{t_3}, where the \textbf{bold} and \underline{underlined} text indicate the optimal and suboptimal results, respectively. It can be seen that our model outperforms all the comparison methods for different evaluation metrics and the specific analysis can be found in \textbf{Appendix \ref{app:E0}}. Additionally, for experiments on semi-supervised node classification, please refer to the \textbf{Appendix \ref{app:E}}.

\begin{table}[!t]
\renewcommand{\arraystretch}{1.0} 
\setlength{\tabcolsep}{2.5pt} 
\centering
\begin{center}
\scalebox{0.95}{
\begin{tabular}{lcccccc}
\hline\multirow{1}{*} & { $N_{neg}$ } & \multicolumn{2}{c} { 50 } & \multicolumn{2}{c}{ 1500 } \\
\hline & Method & Micro-F1 & Macro-F1 & Micro-F1 & Macro-F1 \\
\hline 
\multirow{2}{*}{Cora} & GDCL & \textbf{36.4} & 34.4 & 11.9 & 9.4
\\
 & Ours & $35.6$ & $\mathbf{39.5}$ & $\mathbf{31.8}$ & $\mathbf{36.9}$ \\
\hdashline
\multirow{2}{*}{Citeseer} & GDCL & 32.3 & 31.3 & 4.6 & 3.9
\\
 & Ours & $\mathbf{48.5}$ & $\mathbf{40.3}$ & $\mathbf{48.4}$ & $\mathbf{40.1}$ \\
\hdashline
\multirow{2}{*}{PubMed} & GDCL & 29.0 & 30.0 & 27.1 & 29.2
\\
 & Ours & $\mathbf{39.2}$ & $\mathbf{30.9}$ & $\mathbf{40.4}$ & $\mathbf{32.0}$ \\
\hline
\end{tabular}
}
\caption{The F1 score of graph clustering on three graph datasets between previous contrastive-based SOTA GDCL and our CueGCL. We adopt the F1 score to evaluate the fairness in graph clustering, which can comprehensively reflect the detection accuracy for each cluster.}
\label{t_4}
\end{center}
\end{table}


\begin{table}[!t]
\renewcommand{\arraystretch}{1.0}
\centering
\begin{center}
\scalebox{0.95}{
\begin{tabular}{clccccc}
\hline & Method & Label-free & Cora & Citeseer & PubMed \\
\hline 
\multirow{3}{*}{\textbf{SA}} & Co-training & \XSolidBrush & 73.5 & 62.5 & \textbf{72.7} \\
{} & Union & \XSolidBrush & \underline{75.9} & 66.7 & 70.7 \\
{} & M3S & \XSolidBrush & 75.6 & \underline{70.3} & 70.6 \\
\hline
\textbf{CA} & Ours & \CheckmarkBold & $\mathbf{77.9}$ & $\mathbf{71.4}$ & $\underline{71.2}$ \\
\hline
\end{tabular}
}
\caption{The ACC of community detection on three graph datasets between self-training algorithms and our CueGCL.}
\label{t_3}
\end{center}
\end{table}

\subsection{Ablation Study}
To verify the effect of each module in our CueGCL, we conduct ablation experiments as shown in Table \ref{t_5}.

It is evident that our proposed personalized self-training and alignment module significantly improve the performance of our model for graph clustering. Examining row 1 and row 2 of this table, we can see an increase in performance of 5 to 6 points on the Cora when applying the PeST module. It indicates that this module does allow our model to learn the personalized features of each cluster. Moreover, the change in modularity shown in Figure \ref{g6} further validates the superiority of our PeST module, which leads to more compact cluster structures in embedding space.

\begin{table}[!t]
\renewcommand{\arraystretch}{1.2}
\centering
\begin{center}
\setlength{\tabcolsep}{1.8pt} 
\resizebox{1.02\columnwidth}{!}{
\begin{tabular}{lcccccccccc}
\hline \multirow{2}{*}{ Method } & \multicolumn{2}{c}{ Cora } & \multicolumn{2}{c}{ Citeseer } & \multicolumn{2}{c}{ PubMed } & \multicolumn{2}{c}{ Cornell } & \multicolumn{2}{c}{ Texas } \\
\cline { 2-11 } & ACC & NMI & ACC & NMI & ACC & NMI & ACC & NMI & ACC & NMI\\
\hline w/o $\mathcal{L}_{st}$ \& $\mathcal{L}_{al}$ & $69.9$ & $55.1$ & $69.9$ & $44.4$ & $69.0$ & $35.2$ & $43.7$ & $9.5$ & $44.3$ & $12.4$ \\

w/o $\mathcal{L}_{al}$ & $75.1$ & $59.9$ & $70.8$ & $46.0$ & $70.2$ & $35.2$ & $45.0$ & $11.9$ & $52.9$ & $14.5$ \\

w/o $\mathcal{L}_{st}$ & $75.0$ & $59.5$ & $70.6$ & $46.2$ & $70.1$ & $34.5$ & $44.6$ & $12.3$ & $50.4$ & $13.7$\\

CueGCL & $\mathbf{77.9}$ & $\mathbf{60.5}$ & $\mathbf{71.4}$ & $\mathbf{47.2}$ & $\mathbf{71.2}$ & $\mathbf{36.8}$ & $\mathbf{46.1}$ & $\mathbf{13.6}$ & $\mathbf{56.7}$ & $\mathbf{17.1}$ \\
\hline
\end{tabular}
}
\caption{The effect of each module in our CueGCL for graph clustering on five graph datasets.}
\label{t_5}
\end{center}
\end{table}

\subsection{Hyperparameter Analysis}
To investigate the effect of each hyperparameter on our model, including the number of negative samples $N_{neg}$; the sampling frequency $t$ of Medoids and the dimension $d$ of $MLP$; the trade-off parameters $\gamma_{st}$ and $\gamma_{al}$, we perform experiments and analysis on the Cora dataset. Due to page limitations, the details of the latter two experiments are provided in \textbf{Appendix \ref{A4}}.

\begin{figure}[t]
\centering
\subfigure[ACC]{
\includegraphics[width=0.225\textwidth]{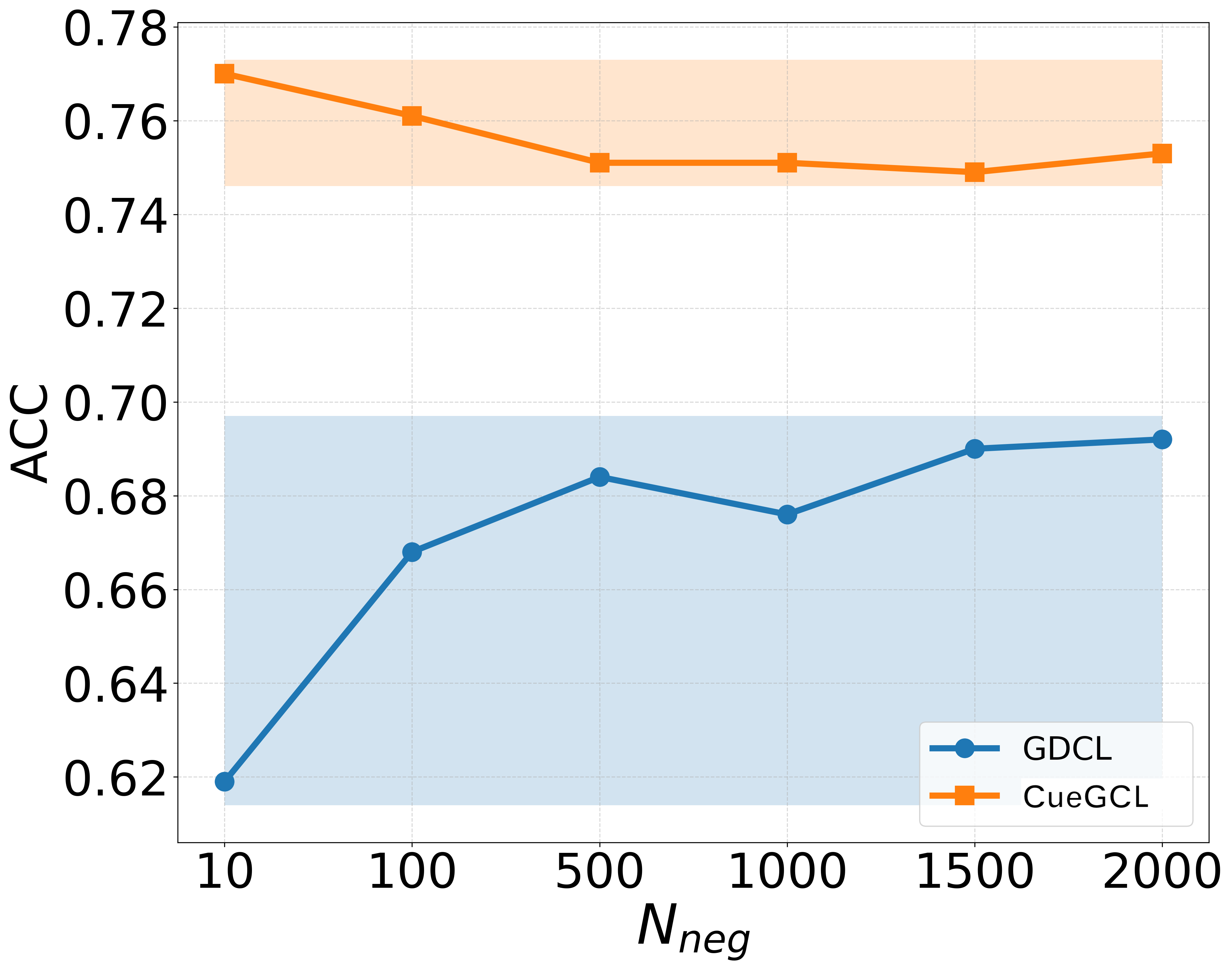}
\label{g7_1}
}
\subfigure[NMI]{
\includegraphics[width=0.225\textwidth]{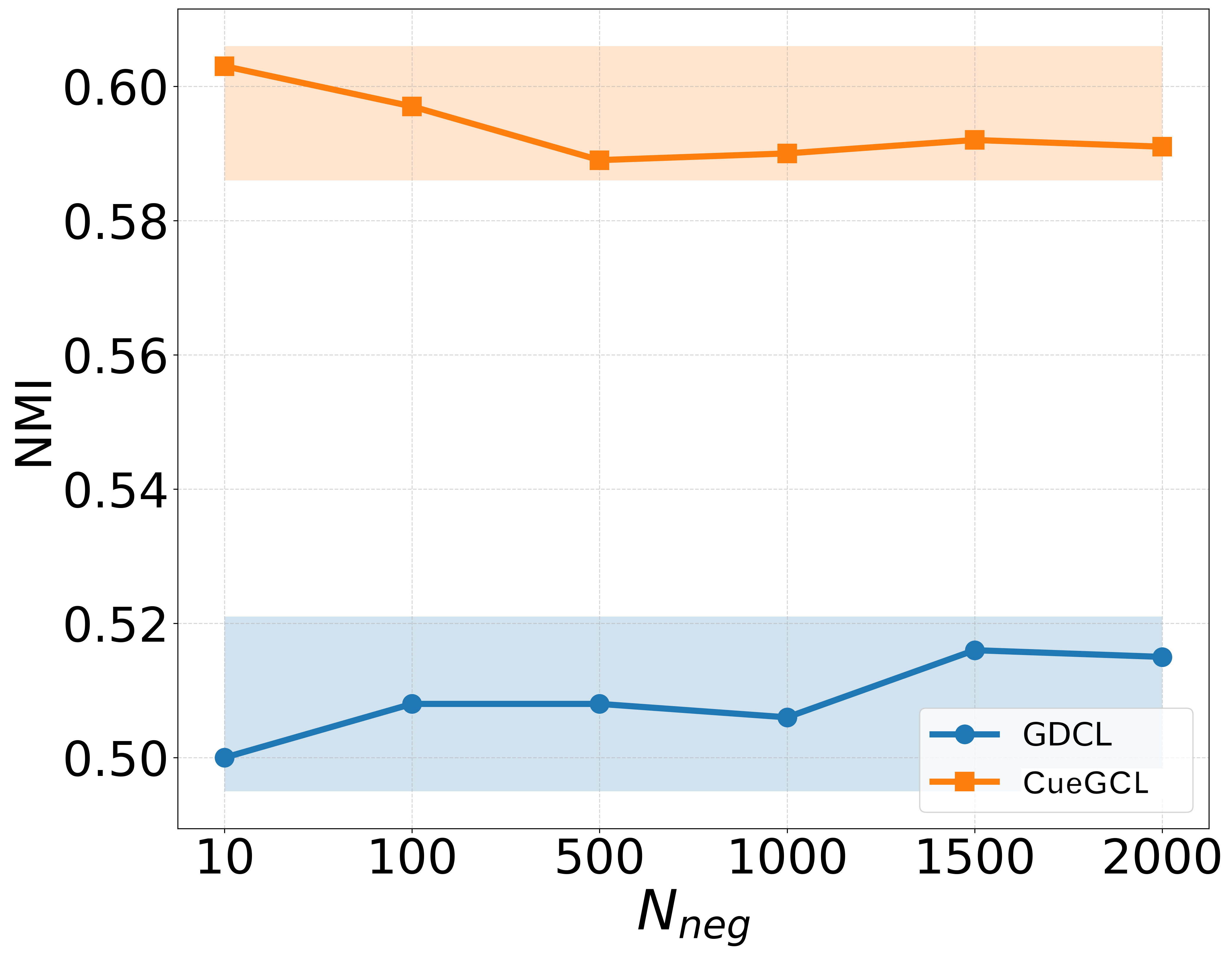}
\label{g7_2}
}
\caption{Graph clustering performance of different negative sample size $N_{neg}$ on Cora. The shaded area indicates the fluctuation range of ACC and NMI.}
\label{fg7}
\end{figure}

\noindent \textbf{The number of negative samples $N_{neg}$. }
Our CueGCL is a graph clustering method based on GCL, so we analyze the effect of $N_{neg}$ and compare it with the GDCL. As can be seen from Figure \ref{fg7}, the performance of GDCL is sensitive to the change of $N_{neg}$. In general, when the number of negative samples is larger, the performance of GDCL is also better, which is in line with the general rule of the contrastive learning model. However, an excessive number of negative samples not only increases the training time, even worse but also leads to the class collision problem. On the contrary, our CueGCL is robust to $N_{neg}$ on both ACC and NMI. In particular, our model achieves optimal results with $N_{neg}=10$ on the Cora dataset. This is attributed to the PeST module, which is equivalent to the contrastive learning of debiased negative samples. Also, we calculate and compare the average accuracy of class information of medoids, with the highest at 86\% for $N_{neg}=10$. 

\section{Conclusion}
In this paper, we propose a cluster-aware graph contrastive learning framework (CueGCL). The personalized self-training strategy can enhance the ability of our model to perceive clusters. In addition, we bring attention to the issue of unfairness in graph clustering when employing existing contrastive learning algorithms. We propose a straightforward yet effective solution to address this concern, marking a novel contribution to the field. By jointly training the three modules of graph contrastive learning, personalized self-training, and aligned graph clustering, we finally obtain an ideal embedding space that is compact within the same cluster and sparse between the distinct clusters. Extensive experiments also demonstrate the superiority of our CueGCL. 

\bibliographystyle{named}
\bibliography{ijcai25}

\clearpage


\begin{singlespace}
\appendix

\noindent To further elaborate on the ideas and results in our paper, this \textbf{supplementary material} provides additional details omitted from the main text. 

\section{Additional Related Work}
\noindent \textbf{Graph Self-Training. }Graph self-training is an important technique in semi-supervised learning that aims to fully exploit the value of data from unlabelled nodes in the graph. The authors in \cite{li2018self-training} propose self-training approaches including Co-training and Union to train graph convolutional network (GCN), which samples predicted labels of nodes with high confidence and adds them to the training set. M3S \cite{sun2020M3S} adopts a multi-stage self-training framework with a deep clustering algorithm. DR-GST \cite{liu2022Robust_ST} considers recovering the distribution of the original labeled graph dataset when introducing the augmented data from self-training. Graph self-training methods can enhance the performance of the model for downstream tasks by improving the generalization of the GCN, but there are fewer self-training mechanisms for unsupervised graph representation learning. Furthermore, due to their limited representation modeling ability, they do not perform as well as our models, even though they use a small number of labels.

\section{Proof of Theorem \ref{Thm3.1}}
\label{A1}
\begin{theorem}[Theorem \ref{Thm3.1} Restated]
An undirected graph $\mathcal{G}$ has $n$ nodes $\mathcal{V}=\left\{v_1, \ldots, v_n\right\}$, and $\boldsymbol{H}_i^{(t)}$ $\left(\boldsymbol{H}_i^{(0)}=\boldsymbol{x}_i\right)$ is the embedding of node $v_i$ after $t$ times message propagating and aggregation. And $\boldsymbol{y}_k \in \mathbb{R}^K$ is an one-hot label vector, where $y^{(k)}=1$ is the $k$-th element of $\boldsymbol{y}_k$. Assume that node $v_i$ belongs to cluster $k$, then we have $\forall \epsilon >0, \exists T \in \mathbb{N}^{+}$, such that $\forall t>T$, then $\left\|\boldsymbol{H}_i^{(t)}-\boldsymbol{y}_k\right\|_2 \leq \epsilon$.
\end{theorem}

\begin{proof}
Given the layer-wise propagation rule of GCN is
$$
\boldsymbol{H}^{(t+1)}=\phi(\tilde{\boldsymbol{D}}^{-\frac{1}{2}} \tilde{\boldsymbol{A}} \tilde{\boldsymbol{D}}^{-\frac{1}{2}} \boldsymbol{H}^{(t)}\boldsymbol{W}),
$$
where $\tilde{\mathrm{\boldsymbol{A}}}={\boldsymbol{A}}+\boldsymbol{I}$ is the adjacency matrix of $\mathcal{G}$ with added self-loops in each node and $\boldsymbol{I}$ is the identity diagonal matrix. $\tilde{\boldsymbol{D}}_{i i}=\sum_j \tilde{\boldsymbol{A}}_{ij}$ and $\boldsymbol{W}$ is weight matrix of GCN. $\boldsymbol{H}^{(t)} \in \mathbb{R}^{n \times d_t}$ is the node embedding matrix in \textit{t-th } layer, $\boldsymbol{H}^{(0)}=\boldsymbol{X} \in \mathbb{R}^{n \times d}$. $\phi$ is the activation function, such as $\operatorname{Relu}$. Here for a easy proof, we assume
that $\phi(x) = x$ and $\boldsymbol{W} = \boldsymbol{I}$. So, We can rewrite the GCN layer as

$$
\boldsymbol{H}^{(t+1)}=\tilde{\boldsymbol{D}}^{-\frac{1}{2}} \tilde{\boldsymbol{A}} \tilde{\boldsymbol{D}}^{-\frac{1}{2}} \boldsymbol{H}^{(t)}=\left(\boldsymbol{I}-\boldsymbol{L}_{\text {sym}}\right) \boldsymbol{H}^{(t)},
$$
where $\boldsymbol{L}_{sym}=\tilde{\boldsymbol{D}}^{-\frac{1}{2}} \tilde{\boldsymbol{L}} \tilde{\boldsymbol{D}}^{-\frac{1}{2}}$ is the symmetrically normalized Laplacian matrix and $\tilde{\boldsymbol{L}}=\tilde{\boldsymbol{D}}-\tilde{\boldsymbol{A}}$. 

Given that each node in the graph $\mathcal{G}$ has a self-loop, it can be concluded that there are no bipartite components in the graph. Consequently, it follows that the eigenvalues of $\boldsymbol{L}_{sym}$ are all contained within the interval [0,2) \cite{chung1997spectral_graph}. As $\boldsymbol{L}_{sym}$ is a real symmetric matrix, it is found to have n eigenvalues, which may include instances of repeated eigenvalues. Additionally, we can derive a set of $n$ orthogonal eigenvectors, denoted as $\{\boldsymbol{e_j}\}_{j=1}^n$. 

Assume that $\mathcal{G}$ has $w$ connected components $\{C_i\}_{i=1}^w$ and define indication vector $\mathbf{1}_i$ as follows:

$$
\mathbf{1}_i^{(j)}=\left\{\begin{array}{l}
1, v_j \in C_i \\
0, v_j \notin C_i
\end{array}\right..
$$
From \cite{von2007tutorial}, we know that the eigenspace of $\boldsymbol{L}_{sym}$ corresponding to the eigenvalue of 0 is spanned by $\left\{\boldsymbol{D}^{-\frac{1}{2}} \mathbf{1}_i\right\}_{i=1}^w$. It is easy to know that if node $v_i$ and $v_j$ in the same connected component, we have

$$
\forall k \in \{1,..,w\}, \quad \mathbf{1}_{k}^{(i)} -  \mathbf{1}_{k}^{(j)} = 0. 
$$
Let $\left(\lambda_1, \ldots, \lambda_n\right)$ denote the eigenvalue of matrix $\boldsymbol{I}-\boldsymbol{L}_{sym}$. Since the eigenvalues of $\boldsymbol{L}_{sym}$ stand in [0,2), we have

$$
-1<\lambda_1<\lambda_2<\cdots<\lambda_{n-w}<\lambda_{n-w+1}=\cdots=\lambda_n=1.
$$

If node $v_i$ and node $v_j$ are in the same connected component, we can rewrite $\boldsymbol{H}_i^{(t)}-\boldsymbol{H}_j^{(t)}$ as

$$
\begin{aligned}
    \begin{array}{l}
    \boldsymbol{H}_i^{(t)}-\boldsymbol{H}_j^{(t)} \\
=\left[\left(\boldsymbol{I}-\boldsymbol{L}_{sym}\right)^{t} \boldsymbol{H}^{(0)}\right]_i-\left[\left(\boldsymbol{I}-\boldsymbol{L}_{sym}\right)^{t} \boldsymbol{H}^{(0)}\right]_j \\
=\left[\left(\boldsymbol{I}-\boldsymbol{L}_{sym}\right)^{t} (\boldsymbol{e}_1,...,\boldsymbol{e}_n)\hat{\boldsymbol{H}}\right]_i-\left[\left(\boldsymbol{I}-\boldsymbol{L}_{sym}\right)^{t} (\boldsymbol{e}_1,...,\boldsymbol{e}_n)\hat{\boldsymbol{H}}\right]_j \\
=\left[\lambda_1^{t}\left(\boldsymbol{e}_1^{(i)}-\boldsymbol{e}_1^{(j)}\right), \ldots, \lambda_{n-w+1}^{t}\times0, \ldots, \lambda_{n}^{t}\times0\right] \hat{\boldsymbol{H}},
\end{array}
\end{aligned}
$$
where $\{\boldsymbol{e_j}\}_{j=1}^n$ are the eigenvectors of $\boldsymbol{L}_{sym}$ (also $\boldsymbol{I}-\boldsymbol{L}_{sym}$) and $\boldsymbol{e}_j^{(i)}$ is the $i$-th element of $\boldsymbol{e}_j$. $\hat{\boldsymbol{H}}$ is the coordinate matrix of $\boldsymbol{H}^{(0)}$ in the $n$-dimensional orthogonal space spanned by eigenvectors of $\boldsymbol{L}_{sym}$. 

We set $\boldsymbol{H}_j^{(t)} = \boldsymbol{M}_k^{(t)}$ and $\boldsymbol{M}_k^{(t)}$ is the embedding of the medoid of the cluster $k$. Thus,

$$
\left\|\boldsymbol{H}_i^{(t)}-\boldsymbol{M}_k^{(t)}\right\|_2=\sqrt{\sum_{j=1}^n\left[\sum_{m=1}^{n-w} \lambda_m^{t}\left(\boldsymbol{e}_m^{(i)}-\boldsymbol{e}_m^{(k)}\right) \hat{\boldsymbol{H}}_{m j}\right]^2}.
$$
Since $-1<\lambda_1<\lambda_2<\cdots<\lambda_{n-w}<1$, we have

$$
\forall \epsilon >0, \exists T_1 \in \mathbb{N}^{+},  \text{s.t. } \forall t>T_1, \text{then} \left\|\boldsymbol{H}_i^{(t)}-\boldsymbol{M}_k^{(t)}\right\|_2 \leq \frac{\epsilon}{2}.
$$

According to the equivalence of cross entropy and L2 norm in the gradient optimization \cite{hinton2015L2}, when we minimize the cross entropy loss  $\mathcal{L}_{st}$ in Eq. \eqref{eq:Lst} from the main text, it is actually equivalent to optimizing the following L2 norm,

$$
\min \left\|\boldsymbol{M}_k^{(t)}-\boldsymbol{y}_k\right\|_2.
$$
So, we have

$$
\forall \epsilon >0, \exists T_2 \in \mathbb{N}^{+},  \text{s.t. } \forall t>T_2, \text{then} \left\|\boldsymbol{M}_k^{(t)}-\boldsymbol{y}_k\right\|_2 \leq \frac{\epsilon}{2}.
$$

In summary, according to the triangle inequality of norm, we have
$$
\begin{aligned}
    & \forall \epsilon >0, \exists T=\max(T_1,T_2),  \text{ s.t. } \forall t>T, \text{then} \\ 
    & \left\|\boldsymbol{H}_i^{(t)}-\boldsymbol{y}_k\right\|_2 \leq 
    \left\|\boldsymbol{H}_i^{(t)}-\boldsymbol{M}_k^{(t)}\right\|_2 + \left\|\boldsymbol{M}_k^{(t)}-\boldsymbol{y}_k\right\|_2 \leq \epsilon. \text{ } 
\end{aligned}
$$
\end{proof}

\section{Algorithm 1}
\label{A5}
\begin{algorithm}[h]
\label{alg1}
\caption{CueGCL}
\begin{algorithmic}
\STATE {\bfseries Input:} Graph $\mathcal{G}$, Iterations $epoch$, Sampling Frequency $t$ 
\STATE {\bfseries Output:} Node Embeddings and Clustering Results
\end{algorithmic}
\begin{algorithmic}[1]
\STATE Initialize model parameters.
\FOR{ $T=1$ \textbf{to} $epoch$ }
    \STATE Encode node embeddings by GCN $g_{\boldsymbol{\theta}}(\cdot)$.
    \IF{$T$ \% $t == 0$}
        \STATE Sample and update the medoid set $M^T$ by Eq. (\ref{eq2}).
        \STATE Update distribution $Q$ and $P$ by Eq. (\ref{eq5}), (\ref{eq7}) respectively.
    \ENDIF
    \STATE Update model parameters by minimizing Eq. (\ref{L}) through SGD.
\ENDFOR
\end{algorithmic}
\end{algorithm}

\begin{table*}[t]
\renewcommand{\arraystretch}{1.5}
\centering
    \scalebox{0.9}{
    \begin{tabular}{|c|c|l|}
    \hline
        Dataset & Backbone & Hyperparameter \\ \hline
        Cora & GCN & epochs: 1000, lr: 5e-5, $hidden_{GCN}$: 512, layers: 1,  $\tau$: 0.5, $N_{neg}$: 10, t: 50, d: 256, $\gamma_{st}$: 0.01, $\gamma_{al}$: 0.001\\ \hline
        Citeseer & GCN & epochs: 1000, lr: 5e-5, $hidden_{GCN}$: 512, layers: 1,  $\tau$: 0.5, $N_{neg}$: 50, t: 150, d: 256, $\gamma_{st}$: 0.0005, $\gamma_{al}$: 0.0001\\ \hline
        PubMed & GCN & epochs: 300, lr: 5e-5, $hidden_{GCN}$: 220, layers: 1, $\tau$: 0.2, $N_{neg}$: 10, t: 30, d: 256, $\gamma_{st}$: 10, $\gamma_{al}$: 0.001 \\ \hline
        Cornell & GCN & epochs: 300, lr: 1.5e-6, $hidden_{GCN}$: 200, layers: 1, $\tau$: 0.5, $N_{neg}$: 4, t: 10, d: 128, $\gamma_{st}$: 10, $\gamma_{al}$: 0.01 \\ \hline
        Texas & GCN & epochs: 50, lr: 8e-6, $hidden_{GCN}$: 512, layers: 1, $\tau$: 0.5, $N_{neg}$: 5, t: 10, d: 512, $\gamma_{st}$: 0.1, $\gamma_{al}$: 0.1 \\ \hline
    \end{tabular}
    }
    \caption{The hyperparameters for our model on five graph datasets.}
    \label{t_7}
\end{table*}

\section{More Details About the Experiment}
\subsection{The Detail of Datasets}
\label{A2}
We conduct experiments on five commonly used datasets for graph learning, using the default dataset splits in PyG\footnote{https://pytorch-geometric.readthedocs.io/}. These datasets primarily include two types of networks and specific details are outlined below.
\begin{itemize}
    \item \textbf{Cora} \cite{sen2008cora}: A citation network with 2708 nodes and 5429 edges. Each node represents a paper in the field of machine learning, and each edge represents a citation relationship between two papers. The feature of a node is represented as a 1433-dimension binary vector, which indicates the presence of the corresponding word. Each node in this graph is classified into one of the following seven classes: Case Based, Genetic Algorithms, Neural Networks, Probabilistic Methods, Reinforcement Learning, Rule Learning, and Theory.
    \item \textbf{Citeseer} \cite{rossi2015citeseer}: A citation network with 3312 nodes and 4732 edges. Each node represents a paper in the field of computer science, and each edge represents a citation relationship between two papers. The feature of a node is represented as a 3703-dimension binary vector. Each node in this graph is classified into one of the following six classes: Agents, AI, DB, IR, ML, and HCI.
    \item \textbf{PubMed} \cite{namata2012pubmed}: A citation network with 19717 nodes and 44338 edges. Each node represents a paper in the field of diabetes, and each edge represents a citation relationship between two papers. The feature of a node is represented as a 500-dimension Term Frequency Inverse Document Frequency (TFIDF) vector. Each node in this graph is classified into one of the following three classes: Diabetes Mellitus, Experimental, Diabetes Mellitus Type 1, and Diabetes Mellitus Type 2.
    \item \textbf{Cornell} and \textbf{Texas} \cite{pei2020cornell}: Two webpage networks with 183 nodes. Each node represents a web page collected from computer science departments of various universities, and each edge represents a hyperlink between two web pages. The feature of a node is represented as a 1703-dimension bag-of-words representation. Each node in both graphs is classified into one of the following five classes: student, project, course, staff, and faculty.
\end{itemize}

\subsection{Hyperparameter Setting}
\label{A3}
Following the suggestions in DEC \cite{xie2016DEC}, we pre-train the encoder of our framework. The important hyperparameters for the graph clustering task are shown in Table \ref{t_7}.

\subsection{Experimental Analysis of Graph Clustering}
\label{app:E0}
\begin{enumerate}
    \item The results in Table \ref{t_2} show that deep learning methods based on GCN for graph clustering are significantly better than traditional clustering algorithms (i.e. K-means, SC). Since GCN makes more reasonable use of graph topology and node properties. Additionally, we observed that the contrastive-based approach (i.e. MVGRL, GDCL, $S^{3}$-CL, and HSAN) generally outperforms the generative-based approach (i.e. VGAE, ARGA, ARVGA, DAEGC, VGAER, and DDGAE). This can be attributed to the fact that the generative-based approach tends to learn feature-level information, whereas the contrastive-based approach learns more node-level information, making it better suitable for downstream graph clustering tasks.
    \item Our proposed CueGCL outperforms all comparison methods including contrastive-based algorithms on all five datasets. For example, on the Cora, our model significantly outperforms the previous SOTA GDCL by 3.2\% and 1.6\% in ACC, and NMI, respectively. It is possible that GDCL could not consider the cluster structure, while our CueGCL explicitly extracts cluster-level information through the integration of PeST and AGC. Furthermore, compared to $S^3$-CL, which utilizes prototype contrastive learning by directly minimizing the distance between prototypes and nodes, our PeST provides a more natural and unbiased approach to cluster structure learning. Based on the experiment results from Figure \ref{g6} and Table \ref{t_2}, it can be concluded that our CueGCL can better solve the \textbf{Challenge \MakeUppercase{\romannumeral 1}} raised in the introduction.
    \item To verify the fairness of contrastive learning methods in graph clustering, we compare our CueGCL with the previous contrastive-based SOTA GDCL in Table \ref{t_4}. As the negative sample size increases and class collision worsens, the fairness of graph clustering based on GDCL receives different degrees of impact on the three datasets. By the way, since the PubMed dataset has only three classes (clusters), GDCL is slightly affected compared to the other two datasets. In contrast, our model reaches significantly higher F1 scores than the GDCL, especially when the negative sample size is large. This indicates our model equally achieves considerable precision when detecting each cluster and solves the problem in \textbf{Challenge \MakeUppercase{\romannumeral 2}} that arises when applying the contrastive framework to graph clustering.
    \item Co-training, Union, and M3S are semi-supervised learning algorithms for self-training. During their training process, we use label rates of 2\%, 3\%, and 0.1\% w.r.t. Cora, Citeseer, and PubMed. It is worth mentioning that the common label rates for these three datasets in node classification tasks are 5.2\%, 3.6\%, and 0.3\%. As shown in Table \ref{t_3}, our model is almost superior to three other methods in terms of ACC when applied to community detection, even though we do not use any label information. This is because our contrastive-based CueGCL fully utilizes the information of abundant unlabeled nodes. Besides, the nodes we sample in the PeST module are more representative of each cluster than those in other self-training methods.
\end{enumerate}

\subsection{Node Classification}
\label{app:E}
To further validate the quality of the embedding space, we conducted node classification experiments based on the standard linear evaluation protocol \cite{velickovic2019DGI,hu2020ogb}. The average accuracy and standard deviations over 10 runs are reported in Table \ref{tab:node_classification}. Specifically, we compared CueGCL with the two categories of state-of-the-art algorithms: (1) Supervised: MLP, GCN \cite{kipf2016GCN}. (2) Unsupervised + Fine-Tuning: VGAE \cite{kipf2016VGAE}, DGI \cite{velickovic2019DGI}, MVGRL \cite{hassani2020MVGRL}, GCA \cite{zhu2021GCA}, BGRL \cite{thakoor2021BGRL}, $S^{3}$-CL \cite{ding2023S3CL} and GraphACL \cite{xiao2024GraphACL}.

We observed that our CueGCL outperforms other methods in terms of average classification performance across the three datasets. This indicates that cluster-level information can also enhance the performance of supervised downstream tasks.

\begin{table}[ht]
\renewcommand{\arraystretch}{1.3}
    \centering
    \scalebox{0.85}{
    \begin{tabular}{lcccc}
        \toprule
        Method & Cora & Citeseer & PubMed & \textit{Avg. Rank} \\
        \midrule
        MLP & 56.11$\pm$0.34 & 56.91$\pm$0.42 & 71.35$\pm$0.05 & 10.0 \\
        GCN & 81.50$\pm$1.30 & 70.30$\pm$0.28 & 78.80$\pm$2.90 & 7.67 \\
        \hline
        VGAE & 76.30$\pm$0.21 & 66.80$\pm$0.23 & 75.80$\pm$0.40 & 9.0 \\
        DGI & 82.30$\pm$0.60 & 71.80$\pm$0.70 & 76.80$\pm$0.60 & 7.0 \\
        MVGRL & 83.03$\pm$0.27 & 72.75$\pm$0.46 & 79.63$\pm$0.38 & 4.33 \\
        GCA & 82.93$\pm$0.42 & 72.19$\pm$0.30 & 80.79$\pm$0.45 & 4.67 \\
        BGRL & 82.70$\pm$0.60 & 71.10$\pm$0.60 & 79.60$\pm$0.50 & 6.33 \\
        $S^{3}$-CL & \underline{84.50$\pm$0.40} & \underline{74.60$\pm$0.40} & 80.80$\pm$0.30 & \underline{2.33} \\
        GraphACL & 84.20$\pm$0.31 & 73.63$\pm$0.22 & \textbf{82.02$\pm$0.15} & \underline{2.33} \\
        \hline
        CueGCL & \textbf{86.42$\pm$0.21} & \textbf{74.88$\pm$0.62} & \underline{81.93$\pm$0.14} & \textbf{1.33} \\
        \bottomrule
    \end{tabular}
    }
    \caption{The performance of node classification with ACC on Cora, Citeseer, and PubMed. The bold and \underline{underlined} text indicates the optimal and suboptimal results, respectively. The \textit{Avg. Rank} denotes the mean value across three datasets for the rank among all methods.}
    \label{tab:node_classification}
\end{table}

\begin{figure}[t]
\centering
\subfigure[ACC]{
\includegraphics[width=0.225\textwidth]{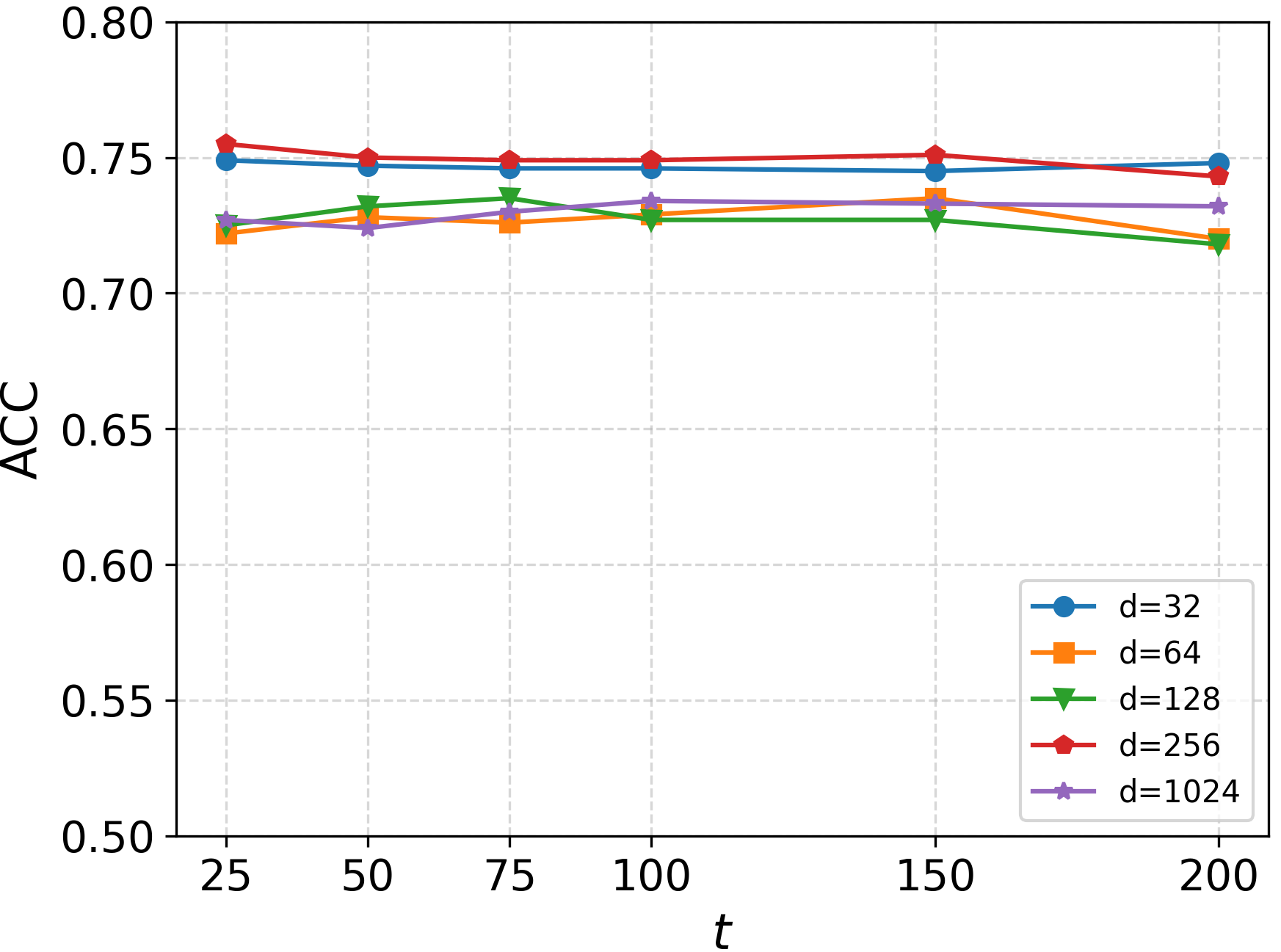}
\label{g8_1}
}
\subfigure[NMI]{
\includegraphics[width=0.225\textwidth]{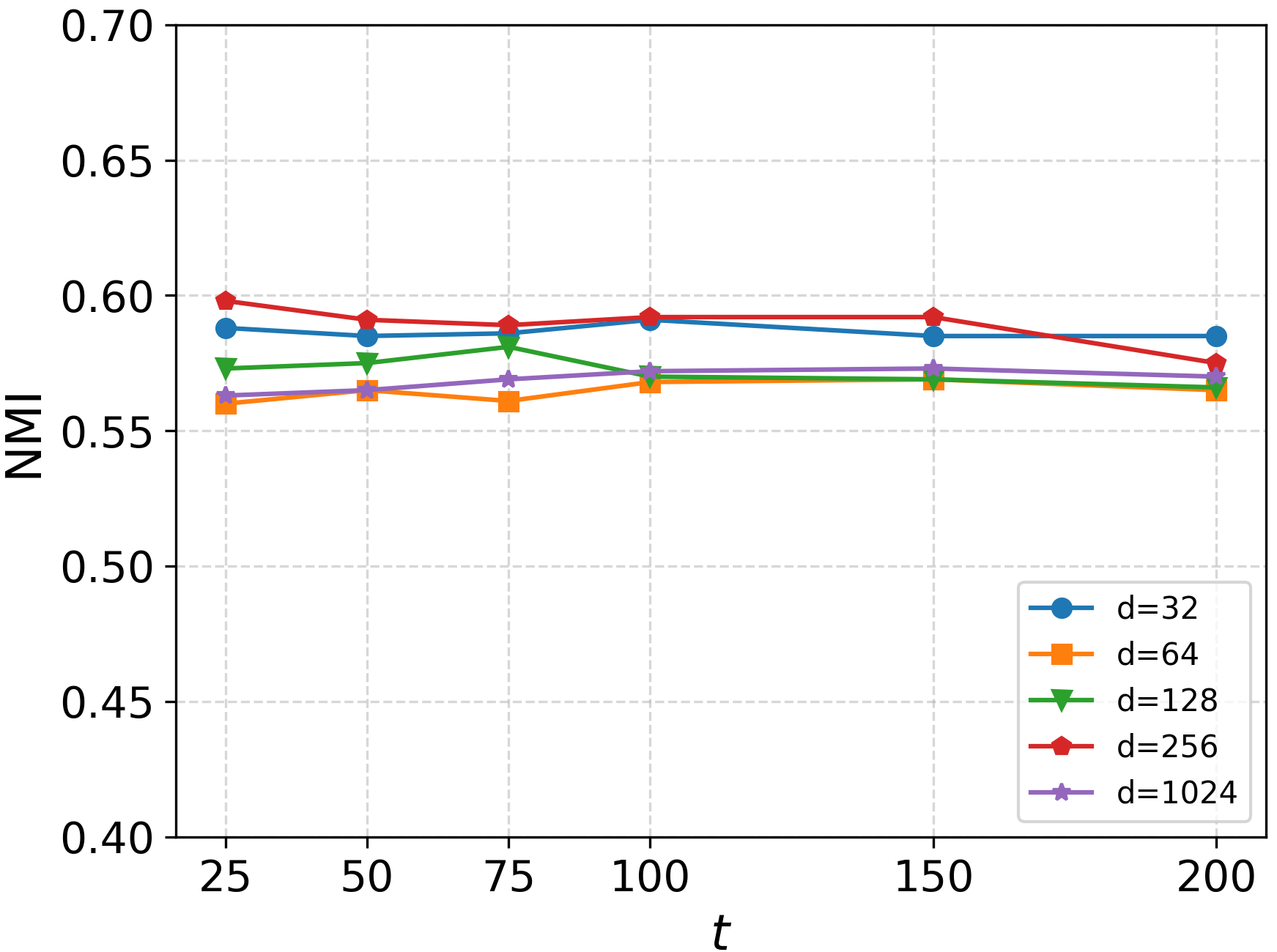}
\label{g8_2}
}
\caption{Clustering performance of different $t$ and $d$ on Cora.}
\label{fg8}
\end{figure}

\begin{table*}[!h]
\renewcommand{\arraystretch}{1.5}
\centering
\begin{center}
\setlength{\tabcolsep}{1.8pt} 
\resizebox{1.5\columnwidth}{!}{
\begin{tabular}{lcccccccccc}
\hline \multirow{2}{*}{ Method } & \multicolumn{2}{c}{ Cora } & \multicolumn{2}{c}{ Citeseer } & \multicolumn{2}{c}{ PubMed } & \multicolumn{2}{c}{ Cornell } & \multicolumn{2}{c}{ Texas } \\
\cline { 2-11 } & ACC & NMI & ACC & NMI & ACC & NMI & ACC & NMI & ACC & NMI\\
\hline Ours with K-means & $74.1$ & $59.1$ & $70.3$ & $45.8$ & $69.6$ & $34.4$ & $42.6$ & $11.3$ & $48.7$ & $12.4$ \\

Ours with K-medoids & $\mathbf{77.9}$ & $\mathbf{60.5}$ & $\mathbf{71.4}$ & $\mathbf{47.2}$ & $\mathbf{71.2}$ & $\mathbf{36.8}$ & $\mathbf{46.1}$ & $\mathbf{13.6}$ & $\mathbf{56.7}$ & $\mathbf{17.1}$ \\


\hline
\end{tabular}
}
\caption{The effect of clustering algorithms in PeST on five graph datasets.}
\label{t_9}
\end{center}
\end{table*}

\section{Additional Hyperparameter Analysis}
\label{A4}
\subsection{The sampling frequency $t$ of Medoids and the dimension $d$ of MLP}
In the PeST module, we sample the medoid vectors of each cluster in every $t$ epochs and generate their prediction labels by using the MLP. The dimension of the hidden layer in MLP is noted by $d$. Then, we conduct graph clustering at $t$ and $d$ from [25, 50, 75, 100, 150, 200] and [32, 64, 128, 256, 1024] respectively, and report the results in Figure \ref{fg8}. As can be seen, the performance of our model remains mostly stable. Furthermore, we can observe that the performance of our CueGCL is slightly higher when $t = 25$ compared to the other $t$. This is due to the fact that more medoid vectors are added to the training set thus allowing our model to learn more information about each cluster in the graph.

\subsection{The trade-off parameters $\gamma_{st}$ and $\gamma_{al}$}
\begin{figure}[t]
\centering
\subfigure[ACC]{
\includegraphics[width=0.225\textwidth]{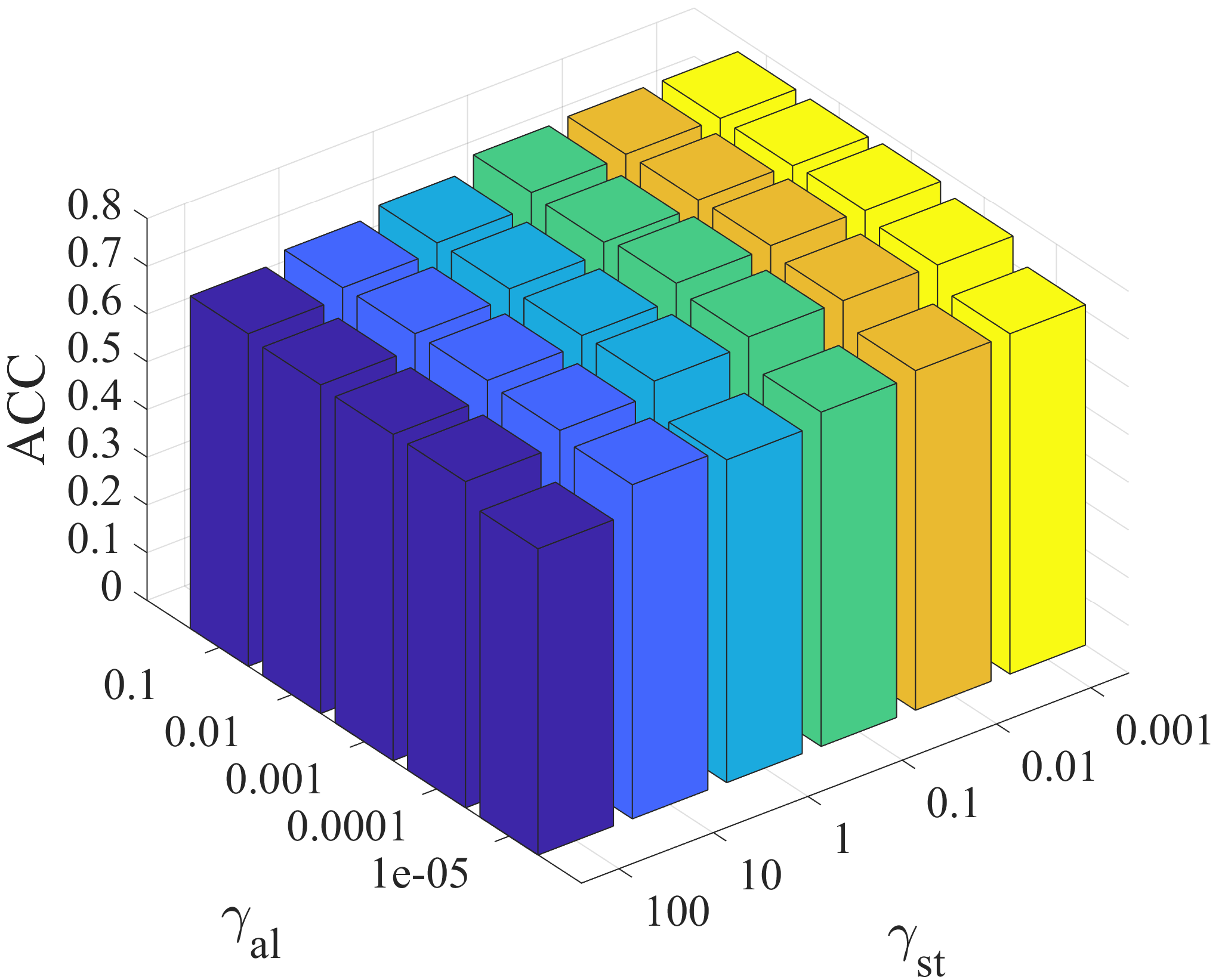}
\label{g6_1}
}
\subfigure[NMI]{
\includegraphics[width=0.225\textwidth]{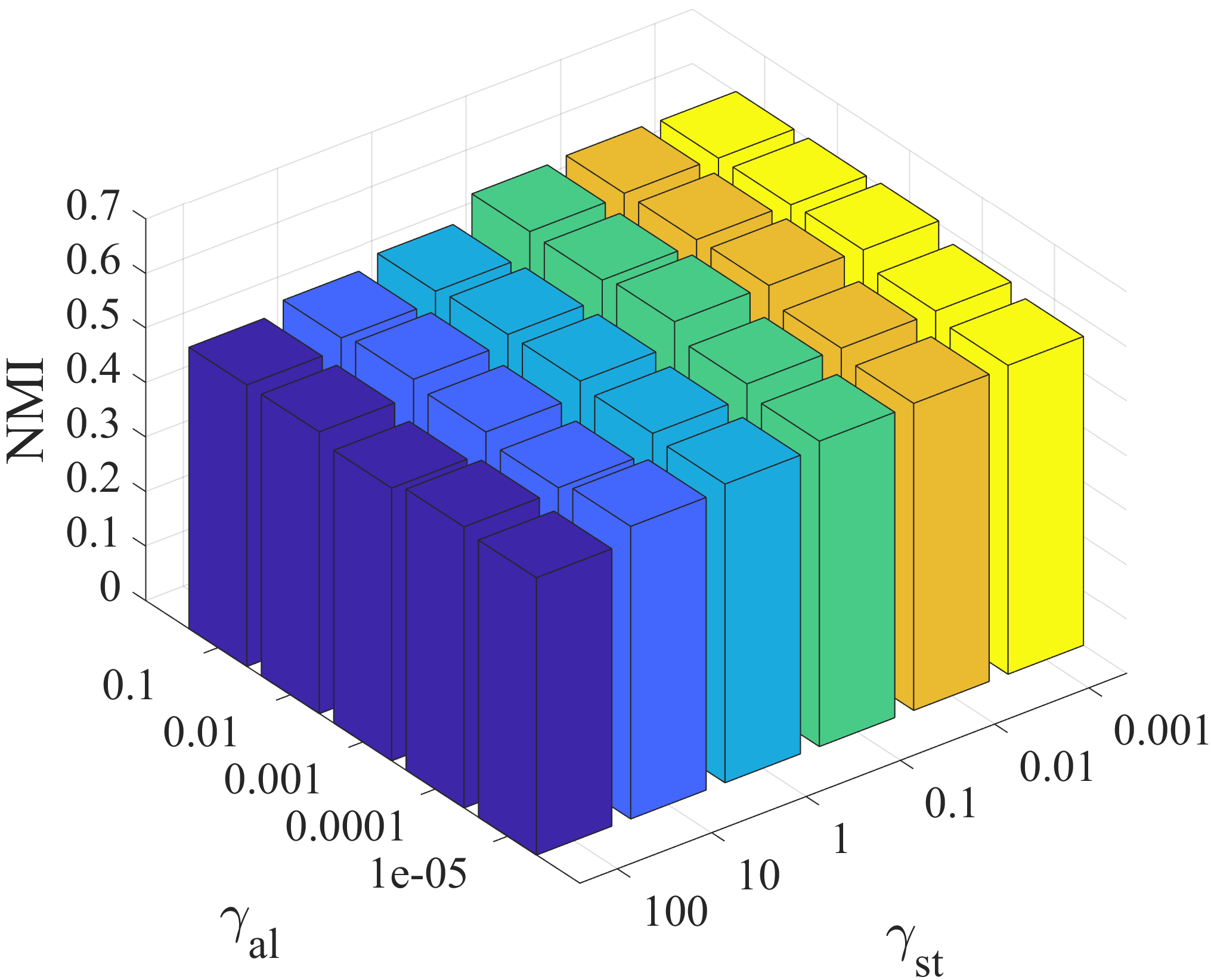}
\label{g6_2}
}
\caption{Clustering performance of different trade-off parameters $\gamma_{st}$ and $\gamma_{al}$ on Cora.}
\label{fg6}
\end{figure}

First, we perform experiments and analyses on the trade-off parameters $\gamma_{st}$ and $\gamma_{al}$, as shown in Figure \ref{fg6}. We can observe that our CueGCL achieves optimal performance on Cora by adopting $\gamma_{st}$ and $\gamma_{al}$ to be 0.01 and 0.001, respectively. For different $\gamma_{st}$, our model is essentially stable on both ACC and NMI. In addition, as $\gamma_{al}$ gradually increases from $10^{-5}$, the performance of our model shows an increasing trend. This also demonstrates the importance of the alignment module in AGC, which indeed improves the generalization ability of our model. Overall, our CueGCL achieves commendable performance with a reasonable combination of trade-off parameters and is robust for both $\gamma_{st}$ and $\gamma_{al}$.

\section{More Details About the PeST}
\label{app:G}
\subsection{On the effectiveness and scalability}
Our proposed PeST is actually more unbiased and efficient contrastive learning of negative samples (as shown in Figure \ref{g4}). The unbiasedness stems from the reliability of the class information. For example, the accuracy of the medoids' class information used for self-training on PubMed remained at \textbf{100\%} throughout the entire training process. In addition, it is evident the time complexity of negative sample learning in general GCL algorithms is $O(n^2)$. While our model's time complexity in negative sample learning is $O(n \times N_{neg})+ O(n \times K) \approx O(n)$, where $N_{neg}$ represents the number of negatives sampled from $\widetilde{\mathcal{N}}_i$, and $K$ represents the number of clusters in the network. Benefiting from the PeST module, only a few negative samples are necessary in our contrastive learning framework. So both $N_{neg}$ and $K$ are much smaller than the number of nodes in the network (i.e. $n$), particularly for large-scale graphs.
\subsection{Why do we use K-medoids in the PeST module instead of other clustering algorithms such as K-means?} 
In our study, we choose the K-medoids algorithm for extracting cluster information and for self-training, based on the following considerations:
\begin{itemize}
    \item \textbf{Motivation Explanation}: As you can see, in lines 268-291 of the paper, we detailed the three main motivations for choosing K-medoids.
    \item \textbf{Theoretical Guarantee}: A significant feature of K-medoids is its use of real nodes from the graph as centroids, providing a theoretical guarantee that the cluster is high-quality from our method. According to lines 745-746 in Appendix \ref{A1}, we need to set $\boldsymbol{H}_j^{(t)} = \boldsymbol{M}_k^{(t)}$, which requires $\boldsymbol{M}_k^{(t)}$ to be the real centroid node (since $H$ represents nodes in the graph), while the centroids obtained by K-means are the mean of nodes within a cluster and may not correspond to any real node.
    \item \textbf{Optimization Loss}: Related work \cite{wu2019feapro} has demonstrated that K-medoids can result in a lower optimization loss compared to other similar clustering methods, such as K-Center, which is effective for various loss functions including cross-entropy we use in Eq. (\ref{eq:Lst}).
\end{itemize}

As shown in Table \ref{t_9}, we replace K-medoids with K-means in our framework, the comparative results show that the performance is better when using K-medoids.

\end{singlespace}

\end{document}